\documentclass[preprint,3p,11pt,times,sort&compress, colorlinks,linkcolor=blue, citecolor=blue]{elsarticle}

\usepackage{amssymb}
\usepackage{amsmath,mathrsfs,lineno}
\usepackage{stfloats}
\usepackage{bm}
\usepackage{graphicx}
\usepackage{booktabs}
\usepackage{float}
\usepackage{caption}
\usepackage{sectsty}
\usepackage{threeparttable}
\usepackage{gensymb}

\usepackage[toc,page]{appendix}
\usepackage[colorlinks,linkcolor=blue,anchorcolor=green,citecolor=blue]{hyperref}
\modulolinenumbers[5]

\journal{xxxxxxx}

\sectionfont{\fontsize{12}{12}\selectfont}
\subsectionfont{\fontsize{11}{11}\selectfont}
\begin{document}

\begin{frontmatter}

\title{Machine learning for predicting fatigue properties of additively manufactured materials}


\author[rvt]{Min Yi}  
\ead{yimin@nuaa.edu.cn}

\author[rvt]{Ming Xue}

\author[syf]{Peihong Cong}

\author[syf]{Yang Song}

\author[syf]{Haiyang Zhang}

\author[res]{Lingfeng Wang}


\author[res]{Liucheng Zhou} 
\ead{happyzlch@163.com}

\author[res]{Yinghong Li} 
\ead{yinghong\_li@126.com}

\author[rvt]{Wanlin Guo} 
\ead{wlguo@nuaa.edu.cn}

\cortext[cor1]{Corresponding author at: College of Aerospace Engineering, NUAA, Nanjing 210016, China}

\cortext[cor1]{Corresponding author}


\address[rvt]{State Key Laboratory of Mechanics and Control for Aerospace Structures \& Institute for Frontier Science \& College of Aerospace Engineering, Nanjing University of Aeronautics and Astronautics (NUAA), Nanjing 210016, China}

\address[syf]{AECC Shenyang Engine Research Institute \& Liaoning Key Laboratory of Impact Dynamics on Aero Engine, Shenyang 110015, China}

\address[res]{Science and Technology on Plasma Dynamics Laboratory, Air Force Engineering University, Xi'an 710038, China}


\begin{abstract}
Fatigue properties of additively manufactured (AM) materials depend on many factors such as AM processing parameter, microstructure, residual stress, surface roughness, porosities, post-treatments, etc. Their evaluation inevitably requires these factors combined as many as possible, thus resulting in low efficiency and high cost. 
In recent years, their assessment by leveraging the power of machine learning (ML) has gained increasing attentions.
Here, we present a comprehensive overview on the state-of-the-art progress of applying ML strategies to predict fatigue properties of AM materials, as well as their dependence on AM processing and post-processing parameters such as laser power, scanning speed, layer height, hatch distance, built direction, post-heat temperature, etc.
A few attempts in employing feedforward neural network (FNN), convolutional neural network (CNN), adaptive network-based fuzzy system (ANFS), support vector machine (SVM) and random forest (RF) to predict fatigue life and RF to predict fatigue crack growth rate are summarized.
The ML models for predicting AM materials' fatigue properties are found intrinsically similar to the commonly used ones, but are modified to involve AM features. 
Finally, an outlook for challenges (i.e., small dataset, multifarious features, overfitting, low interpretability, unable extension from AM material data to structure life) and potential solutions for the ML prediction of AM materials' fatigue properties is provided.
\end{abstract}


\end{frontmatter}

\section{Introduction}
Additive manufacturing (AM), or three-dimensional (3D) printing, allows the production of complex components in a computer-aided design (CAD) model. In contrast to the conventional subtractive manufacturing, AM stands out owing to its merits of large design freedom, near-net shapes, less material waste, etc. 
AM has been demonstrated to be capable of building complex components from different materials such as metals, plastics, ceramics, and polymers~\cite{Park2022,Herzog2016,Zhang2019,Yi2023Modeling}. Thus, AM has great potential to promote the disruptive development of a wide range of industry sectors such as aerospace, biomedical, machinery, automotive, energy, construction, etc.~\cite{Benedetti2021,LI2023456Collaborative,Culmone2019,
ZHU202191Areview,Zhakeyev2017,SHI20201252Anaerospace,DuPlessis2022,
ZHOU20191727Lightweight,Yi2019Computational,Yang20193Dnon}.
However, the AM process is extremely complex, in which various uncertainty parameters could result in printed components with different qualities. In particular, the fatigue properties of AM materials strongly depend on the laser power, scanning speed, layer height, built direction, etc. In order to optimize the performance of AM products, AM experiments are often carried out by setting different processing parameters, which are very expensive and time-consuming. Nevertheless, up to now lots of experiments and theoretical simulations have generated large amounts of data that could be excavated and exploited to serve for the design of AM materials.

To accelerate AM design by utilizing the available data, integrating machine learning (ML) into AM has increasingly become popular in recent years. ML makes it possible to learn from the data automatically and make decisions or predictions without being explicitly programmed~\cite{WANG2020101538Machine,Pham2005,Voulodimos2018}.
Particularly, some ML algorithms, such as support vector machine (SVM) and artificial neural network (ANN), can identify the intricate relationships among nonlinear variables without any specific knowledge or physical model~\cite{CAI2022107580}. Besides, ML methods possess advantages in terms of computation efficiency and prediction accuracy. There have been numerous successful applications of ML for computer vision, image classification, virtual assistants, and autonomous driving~\cite{Mahadevkar2022,Ajani2021,Kaul2022,Sanjana2021,Mozaffari2022}. In AM, ML methods are promising in the determination of mechanical properties of AM materials, and help understand the complex process-structure-property relationships~\cite{Mazhari2021,Rovinelli2018}. For instance, the fatigue properties of AM materials are often influenced by the combination of multiple local factors such as microstructure, porosity, residual stress, surface roughness, etc. It is impossible to develop a unified mathematical model to describe the separate impact and the interaction of these factors. But ML provides a feasible route to harness this issue in AM materials.

\begin{figure*}[!b]
	\centering
	\includegraphics[width=16cm]{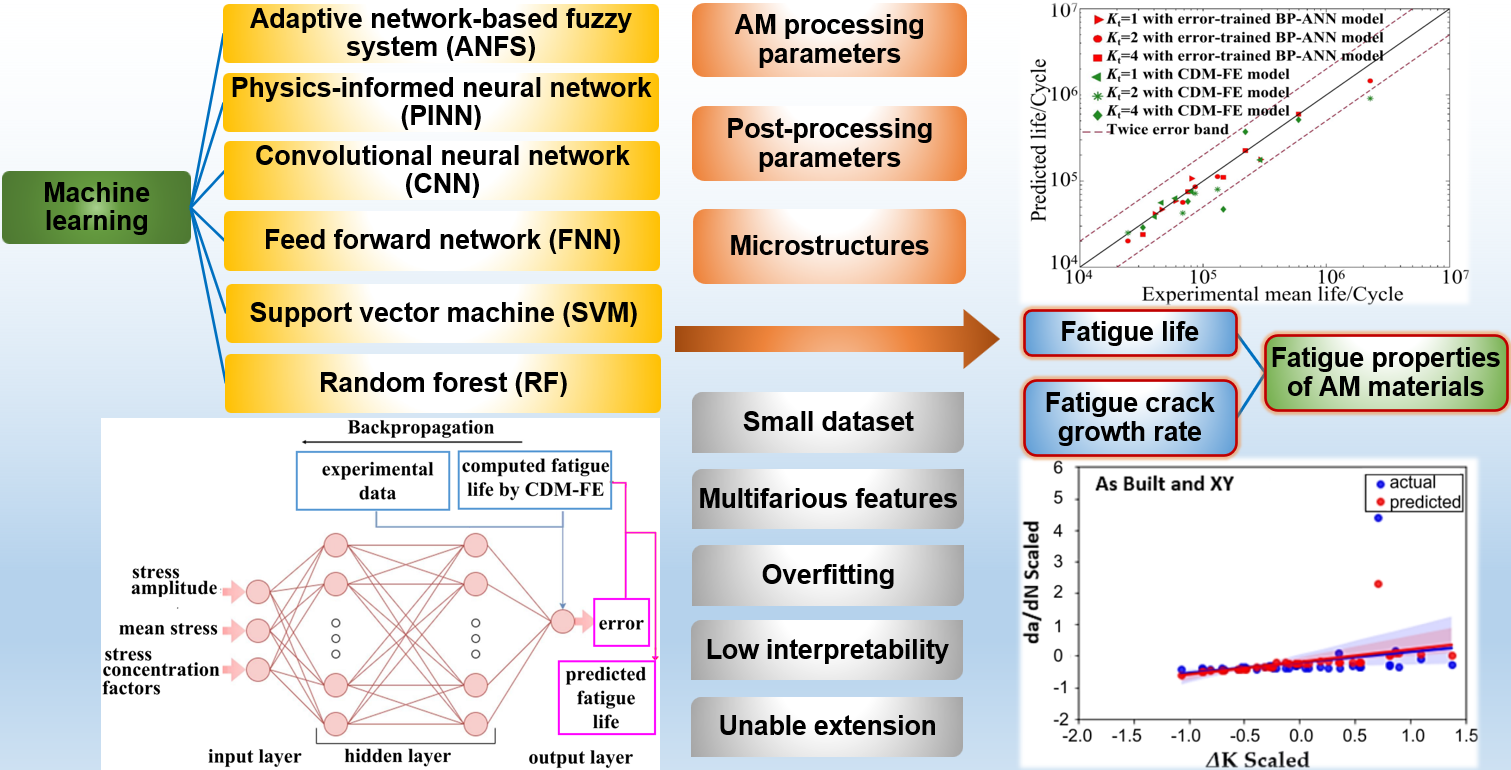}
	\caption{Summary of the state-of-the-art progress of applying ML strategies to the prediction of fatigue properties of AM materials.}
	\label{fig00}
\end{figure*}

Recently, ML has been demonstrated to accelerate the development of AM technology and enable new opportunities in many applications~\cite{WAN20191137,ZHANG2022106808,Zhou13532,GORJI2022106949}. A few review articles in this aspect have become available. For example, some reviews focus on the application of ML to the optimization of AM processing parameters, AM porosity prediction, and defect detection during AM processing~\cite{Nasiri2021,Guo2022,Baumann2018,Meng2020Machine,
Grierson2021Machine}. There is also a review focused on the application of neural network (NN) in fatigue life prediction, fatigue crack behavior, fatigue damage diagnosis, and fatigue strength prediction in subtractive manufactured (not AM) materials~\cite{10.1111/ffe.13640}. A recent review surveys the applications of ML to powder-bed AM technology~\cite{Ladani2021}. Some reviews also discuss the current challenges and future opportunities for algorithmically driven AM processing~\cite{JIN20201541Machine,QI2019721,Kumar2022Machine}. Recent reviews also present ML methods in material and topology design for AM~\cite{WANG2020101538Machine,Raza2022Incorporation,Qin2022}. However, there still lacks a comprehensive overview that specializedly focuses on ML strategies for predicting the fatigue properties of AM materials.

In this Review, we present a systematic overview of the state-of-the-art progress of applying ML strategies to the prediction of fatigue properties of AM materials, as summarized in Fig.~\ref{fig00}. In detail, different ML methods including feedforward neural network, convolutional neural network, adaptive network-based fuzzy system, physics-informed neural network, support vector machine, and random forest for predicting fatigue performance (fatigue life and fatigue crack growth rate) of AM materials are overviewed. For each ML method, its basic knowledge is briefly introduced, followed by its circumstantial applications in predicting fatigue properties of materials (e.g., Ti alloy, stainless steel, Mg alloy, Ni-based alloy, etc.) by AM and different post-processing techniques. Moreover, we provide an outlook for current challenges and potential solutions for the ML prediction of fatigue properties of AM materials.

\section{Feedforward neural network for fatigue life of AM materials}
\begin{figure*}[!b]
	\centering
	\includegraphics[width=16cm]{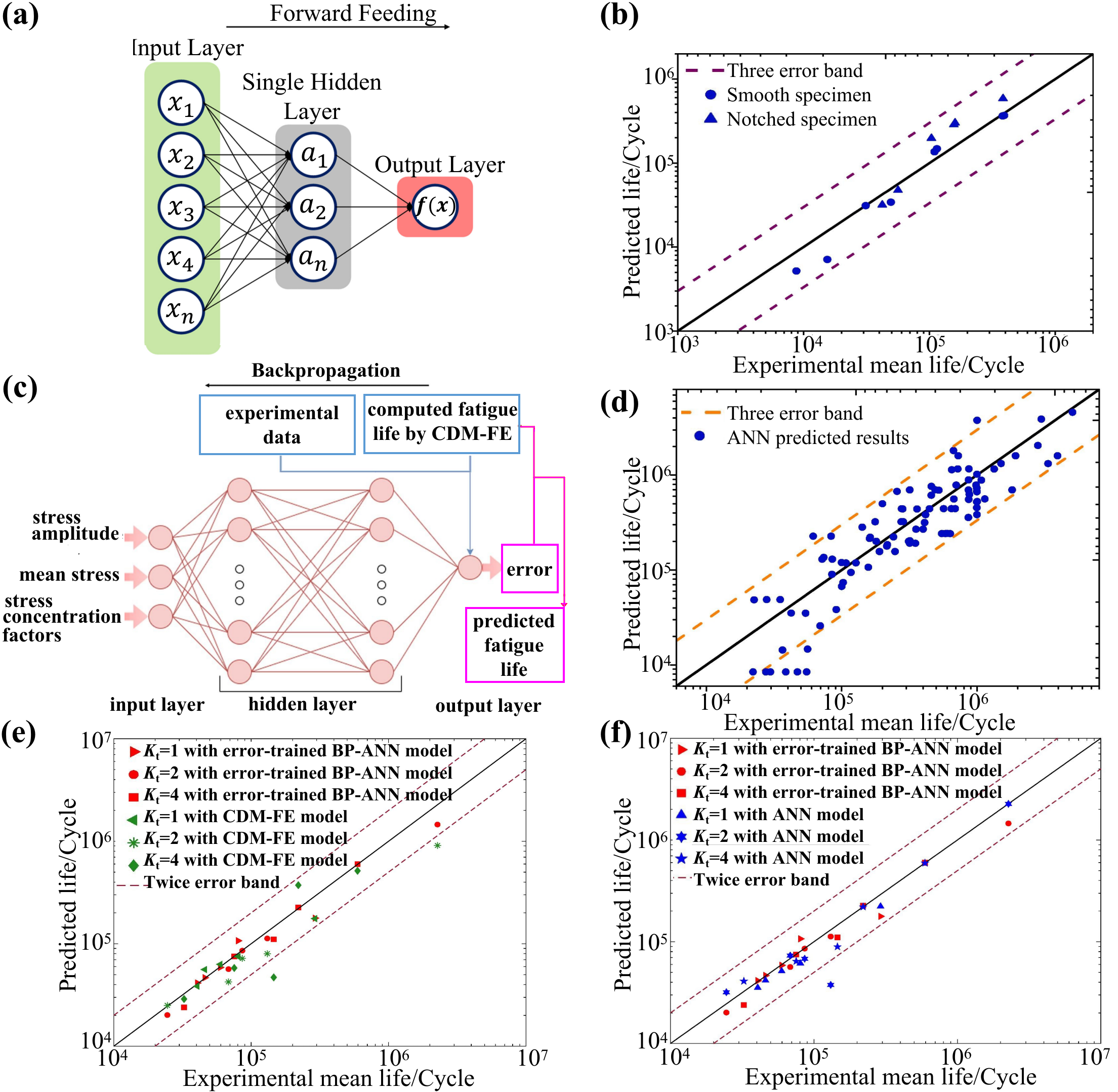}
	\caption{(a) Schematic representation of FNN with single hidden layer~\cite{MUHAMMAD2021102867}; (b) FNN predicted fatigue lives \textit{vs} experimental data for AM processed 300M-AerMet100 steel~\cite{ZHAN2022108352}; (c) Schematics for the error-trained BP-ANN~\cite{LIU2022106836}; (d) Variation of FNN predicted fatigue life \textit{vs} experimental data for AM SS316L \cite{ZHAN2021106089};  Variations of fatigue lives for LC4 aluminum alloys predicted by (e) error-trained BP-ANN model and CDM-FE method with the experimental results, and (f) error-trained BP-ANN model and ANN model trained only with the experimental data ~\cite{LIU2022106836}. }
	\label{fig01}
\end{figure*}

As shown in Fig.~\ref{fig01}(a), there are three types of layers (input, hidden, and output layers) in a typical artificial neural network (ANN)~\cite{SCHMIDHUBER201585}. The information passes from the input layer through the hidden layer to the output layer. Each layer contains a number of neurons and the calculations within a neuron are performed as~\cite{XU201018}
\begin{gather}
	z_k^{(l)}=b_k^{(l-1)}+\sum_{j=1}^{p_{l-1}}w_{kj}^{(l-1)}a_j^{(l-1)}\quad l=2,...,L,  \\
	a_k^{(l)}=g_k^{(l)}\left(z_k^{(l)}\right)\quad k=1,2,...,p_l,
\end{gather}
where $l$ is the layer index with $l=1$ as the input layer and $l=L$ as the output layer, and $p_l$ is the number of neurons at the $l$-th layer. For the input layer, one has
\begin{equation}
	a_k^{(1)}=x_k\quad k=1,...,p_1 .
\end{equation} 
$g_k^{(l)}$ is the activation function related to the $k$-th neuron in layer $l$, which could be sigmoid, rectified linear unit (ReLU), and hyperbolic tangent functions~\cite{MUHAMMAD2021102867}. The intercepts $b_k$ are biases and coefficients $w_{kj}$ are weights. The loss function is a measure of the distance between the ANN outputs and the desired values. The biases and weights are updated by the training process to make the loss function minimized~\cite{SCHMIDHUBER201585}.

For a feedforward neural network (FNN) in Fig.~\ref{fig01}(a), there are two phases, i.e., forward and backward phase~\cite{10.1111/ffe.13640,LI20031861}. In forward phase, the neurons are sequentially activated from the input to output layer to calculate an output signal. In backward phase, the error defined as difference between the output signal and the true value is propagated backward in the network to adjust the connection weights between neurons and reduce future errors. For changing the weights, optimization techniques such as stochastic gradient descent~\cite{lan2017conditional} and Adam algorithm~\cite{LIU2019129} are used.

FNN has been used to predict the fatigue life of AM materials. In detail, Zhan et al.~\cite{ZHAN2022108352} investigate the fatigue performance of AM processed 300M-AerMet100 steel by combining experimental data, numerical simulations, and machine learning. As a first step, they carry out experiments to obtain fatigue curves as calibration and determine the parameters in theoretical models. Then, fatigue models derived from continuum damage mechanics (CDM) are numerically implemented and a large number of simulations are performed to generate sufficient training data for FNN.
In their study, the training data are comprised of stress concentration factor $K_t$, stress ratio $R$, and maximum nominal stress $\sigma_{\text{max}}$. The FNN output is the fatigue life $N_f$. About 200 groups of data with different fatigue loads are employed to train the FNN. For evaluating the FNN performance, the coefficient of determination ($R^2$) and the mean squared logarithmic error ($MSLE$) are utilized, i.e.,
\begin{equation}
R^2\left(N,N^{pre}\right)=1-\frac{\sum_{k=1}^{n}\left(N_k-N_k^{pre}\right)^2}{\sum_{k=1}^{n}\left(N_k-N_m\right)^2}
\end{equation}
and
\begin{equation} \label{msleEq}
	MSLE\left(N,N^{pre}\right)=\frac{1}{n}\sum_{k=1}^{n}\left[\ln\left(1+N_k\right)-\ln\left(1+N_k^{pre}\right)\right]^2,
\end{equation}
in which $N_k$ is the $k$-th experimental fatigue life, $N^{pre}_k$ the $k$-th predicted fatigue life, and $N_m$ the median value of experimental fatigue lives.

All the predicted results from the CDM-FNN method are found to locate within the three-error band in Fig.~\ref{fig01}(b) ($R^2=0.65$, $MSLE=0.18$). To improve the FNN accuracy, the influence of the number of hidden layers ($n_\text{hl}$), the number of neurons ($n_\text{n}$) and the number of training datasets ($n_\text{td}$) on the predicted results is examined. It is found that the FNN performance could be improved by increasing $n_\text{n}$ and $n_\text{td}$, and $n_\text{hl}$ larger than 2 has no notable contribution to the improvement of prediction performance.

Similarly, the influence of AM processing parameters on the fatigue properties of AM alloy is examined~\cite{ZHAN2021106089}. In detail, the CDM-FNN method is applied to compute the fatigue behavior of three AM aerospace alloys including SS316L, Ti6Al4V and AlSi10Mg. The input data include four AM processing parameters (laser power $P$, scanning speed $v$, hatch space $h$, powder layer thickness $t$) and two fatigue loads parameters (maximum stress $\sigma_{\text{max}}$, stress ratio $R$). The output is fatigue life $N_f$. Two metrics are employed to reasonably assess the predicted performances of FNN. In addition to the mean squared logarithmic error $MSLE$ in Eq. (\ref{msleEq}), the mean absolute error $MAE$ is also adopted, i.e.,
\begin{equation}
	MAE\left(N,N^{pre}\right)=\frac{1}{n}\sum_{k=1}^{n}|N_k-N_k^{pre}| .
\end{equation}

For AM SS316L, both the FNN predicted and experimental fatigue lives are presented in Fig.~\ref{fig01}(d). 99 sets of fatigue lives are split into the training data, validation data and test data. It is found that the results from both the trained and validated data are located within the three-error band, but eight of the tested data results are beyond the error band. For AM Ti6Al4V and AM AlSi10Mg, the number of tested data results beyond the error band becomes five and four, respectively. According to the parametric studies on FNN model for life prediction of AM alloys, it is suggested that the FNN model with over 3 hidden layers and with over 20 neurons could effectively improve the prediction accuracy~\cite{ZHAN2021106089}.

Similar approach using the error-trained back propagation artificial neural network~(BP-ANN) is proposed to predict the high cycle fatigue life of LC4 aluminum alloys~\cite{LIU2022106836}. As shown in Fig.~\ref{fig01}(c), the fatigue life is not directly used as the ANN output. In contrast, the relative error ($\delta$) between the experimental life ($N_{exp}$) and computed fatigue lives ($N_{num}$) by CDM method is chosen as the training target, i.e.,
\begin{equation}
	\delta=\frac{N_{num}-N_{exp}}{N_{exp}}\times 100\%.
\end{equation}
$\delta$ acts as an indicator to adjust the numerical results. The final predicted fatigue lives ($\tilde{N}$) are defined as
\begin{equation}
	\tilde{N}=\frac{N_{num}}{1+\delta}.
\end{equation}
To train this BP-ANN model, a total of 106 groups of experimental fatigue lives and CDM computed fatigue lives by finite element (FE) method with different stress concentration factors ($K_t$), stress mean values ($\sigma_m$) and stress amplitudes ($\sigma_a$) for LC4 parts are obtained. $K_t$, $\sigma_m$ and $\sigma_a$ are the input of ANN model.
Fig.~\ref{fig01}(e) and (f) present the fatigue lives from the CDM-FEM method informed with the experimental results, the error-trained BP-ANN model, and the ANN model trained only with the experimental data. The predicted fatigue lives by the error-trained BP-ANN model are found to lie within the twice-error band. Only two results predicted by the CDM-FE model are beyond the twice-error band. The CDM-FE method in Fig.~\ref{fig01}(e) is revealed less efficacious for the prediction of high cycle fatigue life when the life is beyond $10^5$. The error-trained BP-ANN model possesses a correlation coefficient closer to 1 and smaller prediction error, indicating that it is superior than the other two methods. Furthermore, the ANN models with different network structures, i.e., different hidden layers and different neurons at each hidden layer, are utilized to predict the fatigue lives of three data points from the test set. It is found that the error-trained BP-ANN model could maintain relatively stable prediction performance, but the prediction accuracy cannot be guaranteed and there exits obvious volatility in the predicted results from the ANN model trained only with the experimental data~\cite{LIU2022106836}.

\begin{figure*}[!b]
	\centering
	\includegraphics[width=16cm]{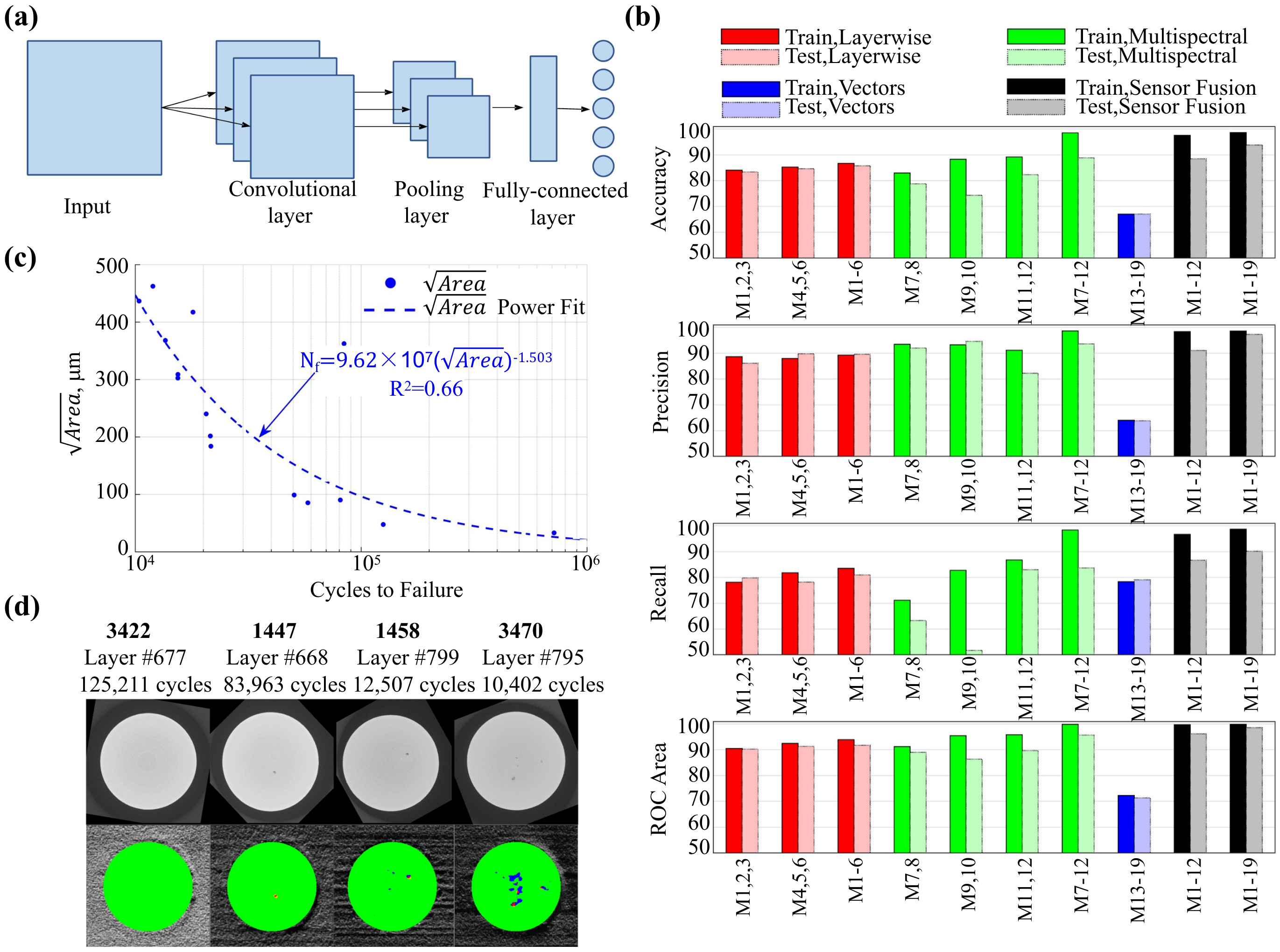}
	\caption{(a) The schematic diagram of a convolution neural network~\cite{10.1111/ffe.13640}; (b) Validation and testing performance of trained CNN classifiers via different merged data modalities~\cite{SNOW2022117476}; (c) Fatigue life \textit{vs} Murakami factor ($\sqrt{area}$)~\cite{SNOW2022117476}; (d) Fatigue life, X-ray tomography scan, and CNN classification results for four fatigue parts from the training and testing data sets~\cite{SNOW2022117476}. For the ML responses, TP voxels are marked by red, TN voxels by green, FP voxels by blue, and FN voxels by yellow.}
	\label{fig03}
\end{figure*}

\section{Convolutional neural network for fatigue life of AM materials}
A convolutional neural network (CNN) is an assembly of neurons that contain input layers, convolution layers, pooling layers, and fully connected layers~\cite{10.1111/ffe.13640}, as shown in Fig.~\ref{fig03}(a).
Operations of convolution and activation are carried out in the convolution layers.
In convolution as a linear operation, multiplication between an array of weights and an array of input data is carried out. This multiplication is called a kernel. 
The kernel usually possesses a specific size, e.g., $3\times3$ or $5\times5$~\cite{Vaz2021}. A convolution layer with $n$ kernels can detect $n$ local features that give rise to the formation of $n$ feature maps~\cite{7426826}. One could create a feature map from the input layer to the hidden layer by a filter convolving over the image.
The spatial size of original image could be reduced and thus each feature map could be condensed by the pooling layer~\cite{LeCun2015}.

Snow et al.~\cite{SNOW2022117476} have trained a CNN model to study the correlation between the \textit{in-situ} sensor data during laser powder bed fusion (LPBF), the internal part quality, and the fatigue performance. 
In their work, two builds (named as C003 and C001) are fabricated by LPBF of Ti6Al4V powders. 30 parts from C003 are selected to produce a training data set for CNN. Additional 9 parts from C001 are chosen for the testing and evaluation of CNN classifier generalizability. For the testing of high-cycle fatigue, 15 (10 from C003 and 5 from C001) fatigue parts are picked. The fractographic surface is analysed to identify the failure origin of each part. The maximum depth from the sample surface and the cross-section area are recorded. Three types of monitoring data during AM processing, i.e., layerwise images (M1 to M6), multi-spectral emissions (M7 to M12), and laser scanning vector data (M13 to M19), are gathered and utilized for the CNN. 

The standard metrics for a binary classification are applied to assess CNN performance. In detail, the $2 \times 2$ confusion matrix possesses four elements that represent true positives ($TPs$), true negatives ($TNs$), false positives ($FPs$), and false negatives ($FNs$). Specifically, one could adopt the following metrics
\begin{gather}
	accuracy=\frac{TPs+TNs}{TPs+FPs+TNs+FNs}, \\
	precision=\frac{TPs}{TPs+FPs}, \\
	recall=\frac{TPs}{TPs+FNs}.
\end{gather}
The receiver operating characteristic (ROC) curves are utilized as another important metric to describe the trade-off between the true and false positive rates when the decision threshold is altered.

As shown in Fig.~\ref{fig03}(b), it is found that spectral process emissions sensors (M7 to M12, green) and layerwise images (M1 to M6, red) provide the most information. In contrast, the commanded machine vectors (M13 to M19, blue) provide little useful information. If the CNN training is performed by using the fused data from all 19 data modalities (M1 to M19, black and gray), the best classifier could be achieved to reach a ROC curve area and a validation accuracy of 99.7\% and 97.4\%, respectively. When the independently generated testing data are used to train CNN, the obtained value for ROC curve area and validation accuracy is reduced to 98.9\% and 93.9\%, respectively.
Then the trained CNN is applied to the case of 200 layers. These layers in total contain the gauge regions of four selected fatigue parts that cover a wide range of fatigue lives from the testing and training data. As found in Fig.~\ref{fig03}(d), the classifier correctly predicts that there is no flaw of a spherical equivalent diameter larger than 200~$\mu$m on the part's failure-initiation layer. In this case, the part possesses a fatigue life of 125,211 cycles. From the Murakami methodology in Fig.~\ref{fig03}(c), it can be seen that the classifier could predict a fatigue life greater than 33,000 cycles (corresponding to $\sqrt{area}\sim 200$~$\mu$m). For the case of other three fatigue parts in Fig.~\ref{fig03}(d) with large flaw densities, the fatigue life is remarkably reduced. This fact indicates a correlation between the fatigue life and the CNN classification response.

\section{Adaptive network-based fuzzy system for fatigue life of AM materials}
\begin{figure*}[!b]
	\centering
	\includegraphics[width=16cm]{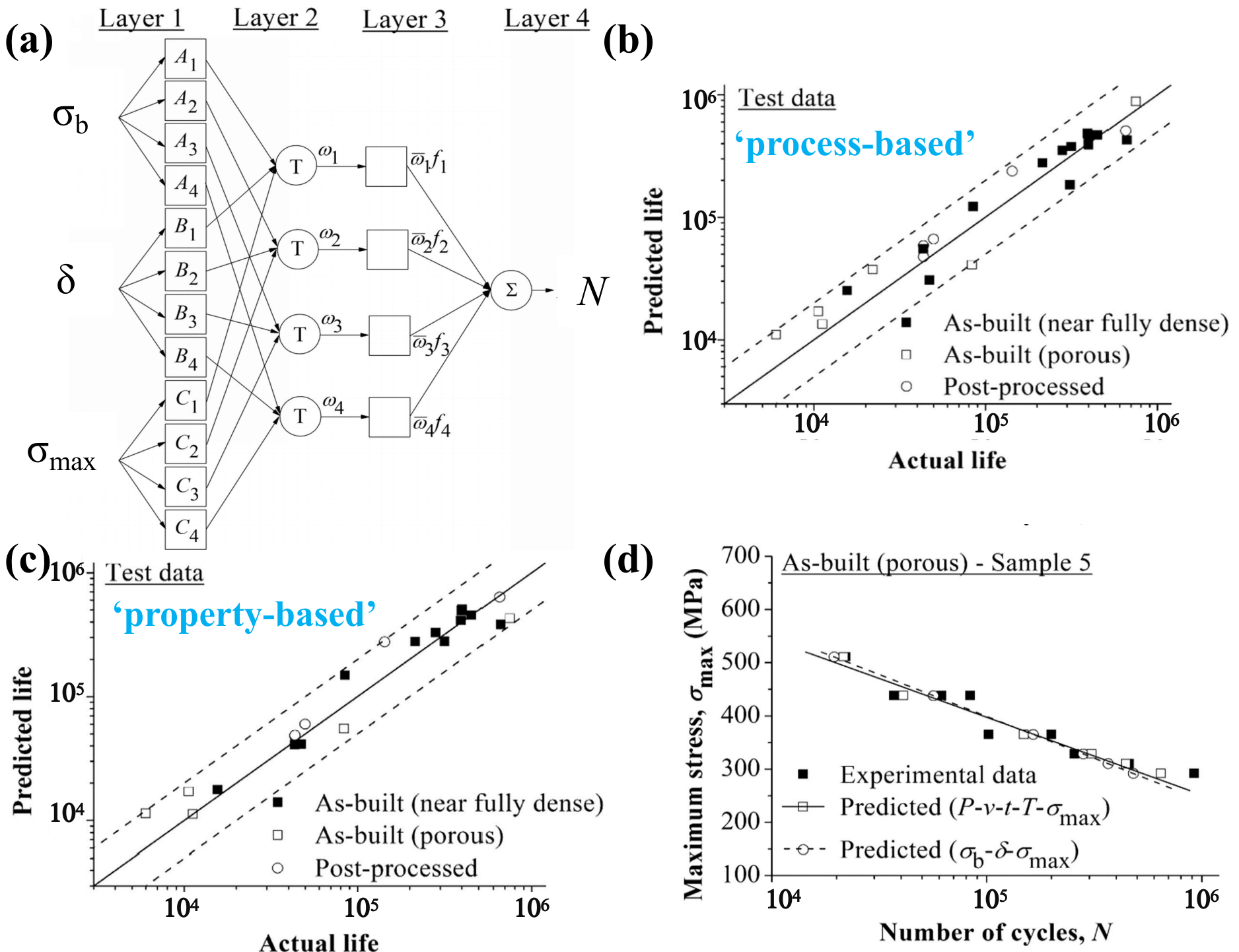}
	\caption{(a) Structure of the adaptive network-based fuzzy system (ANFS)~\cite{ZHANG2019105194}; Experimental and predicted fatigue life by (b) process-based model and (c) property-based model, with dotted lines representing the two-error band~\cite{ZHANG2019105194}; (d) Experimental and predicted S--N curves for selected samples~\cite{ZHANG2019105194}.}
	\label{fig06}
\end{figure*}

Combining the architecture of Takagi--Sugeno fuzzy inference systems with the supervised learning ability from radial basis function neural network~\cite{LIMAJUNIOR2020106191}, Jang~\cite{256541} proposes the adaptive network-based fuzzy system (ANFS).
In contrast to FNN models, in ANFS there are no synaptic weights for connections between the nodes from different layers~\cite{Aengchuan2018}. An ANFS possesses a neural network structure that is constructed by using a set of 'if-then' rules, e.g., 'if $x$ is $A$, then $y$ is $B$'~\cite{TAKAGI198355}.
As shown in Fig.~\ref{fig06}(a), there are four layers in an ANFS. The circle nodes are fixed nodes and the square squares are adaptive nodes. The outputs of adaptive nodes depend on parameters within each node, which are tuned in the training stage by using a set of sample values. The ANFS model in Fig.~\ref{fig06}(a) contains input parameters $\sigma_b$, $\delta$ and $\sigma_{\text{max}}$ that are partitioned into classes $A_j, B_k$ and $C_l$, respectively~\cite{ZHANG2019105194}.
The output of Layer 1 can be calculated by a membership functions $\mu$, i.e.,
\begin{gather}
	O^1_j=\mu_{Aj}\left(\sigma_b\right)\quad j=1,2,3,4  \\
	O^1_k=\mu_{Bk}\left(\delta\right)\quad k=1,2,3,4    \\
	O^1_l=\mu_{Cl}\left(\sigma_{\text{max}}\right) \quad  l=1,2,3,4,
\end{gather}
where membership functions could be triangular, trapezoidal, Gaussian or bell-shaped functions. Layer 2 performs the T-norm operation in which membership values of each fuzzy rule are multiplied and thus generates the output as
\begin{equation}
	O^2_i=\omega_i=\mu_{Aj}\left(\sigma_b\right)\times\mu_{Bk}\left(\delta\right)\times\mu_{Cl}\left(\sigma_{\text{max}}\right)\quad i=1,2,3,4.
\end{equation}
The output $\omega_i$ is the firing strength or known as 'degree of fulfilment' of a particular rule.
The output of a node in Layer 3 is the product of the consequent function $f_i$ with the normalized firing strength ($\bar{\omega_i}$, the ratio of the $i$-th rule's firing strength to the summation of firing strengths of all the rules), i.e.,
\begin{equation}
	O^3_i=\bar{\omega_i} f_i.
\end{equation}
Finally, the summation of all incoming signals generates the output of the last layer, i.e.,
\begin{equation}
	O^4=\sum_{i}\bar{\omega_i}f_i
\end{equation}

In the prediction of high cycle fatigue life of LPBF-AM 316L stainless steel, Zhang et al.~\cite{ZHANG2019105194} examines the capability of ANFS method.
The dataset, which consists of 139 fatigue life data for samples subjected to various processing parameters (laser power $P$, scanning speed $v$, and layer thickness $t$), different post-processing treatments (annealing and hot isostatic pressing), and different mechanical properties (maximum applied cyclic stress $\sigma_{\text{max}}$, ultimate tensile stress $\sigma_b$, and elongation to failure $\delta$), is constructed to realize a complex nonlinear input-output environment. Specifically, two ANFS models with different input variables, i.e., process-based model and property-based model, are implemented to predict the fatigue life of LPBF-AM 316L stainless steel.

The process-based model chooses $P, v, t, T$ and $\sigma_{\text{max}}$ as inputs. The input $T$ is set to possess levels 1, 2, 3 and 4 for representing the as-built, low-temperature annealing, high-temperature annealing and hot isostatic pressing conditions, respectively. On the contrary, the property-based model takes $\sigma_{\text{max}}, \sigma_b$ and $\delta$ as inputs. After the evaluation of the performance of ANFS model, the fuzzy rules for the process-based model and the property-based model are found to be nine and four, respectively.
Fig.~\ref{fig06}(b) and (c) compare the experimental fatigue life and the predicated ones by these two models. The predicted fatigue lives are generally within a tow-error band of the experimental data, even for the porous samples that present the greatest errors. Fig.~\ref{fig06}(d) presents the predicted and experimental result in S--N plots for the selected samples. The predicted results by both models at each loading condition are found to lie within the scatter band of experimental data. This indicates that both models are capable of accounting for variations in the dataset due to the fatigue scatter. In addition, a direct application of ANFS model to data from literature results in a range of prediction accuracy owing to the variability of the reported data. It is suggested that the performance of ANFS model could be further improved by retraining the model with the literature data added into the dataset.

  
\section{Physics-informed neural network for fatigue life of AM materials}
\begin{figure*}[!b]
	\centering
	\includegraphics[width=16cm]{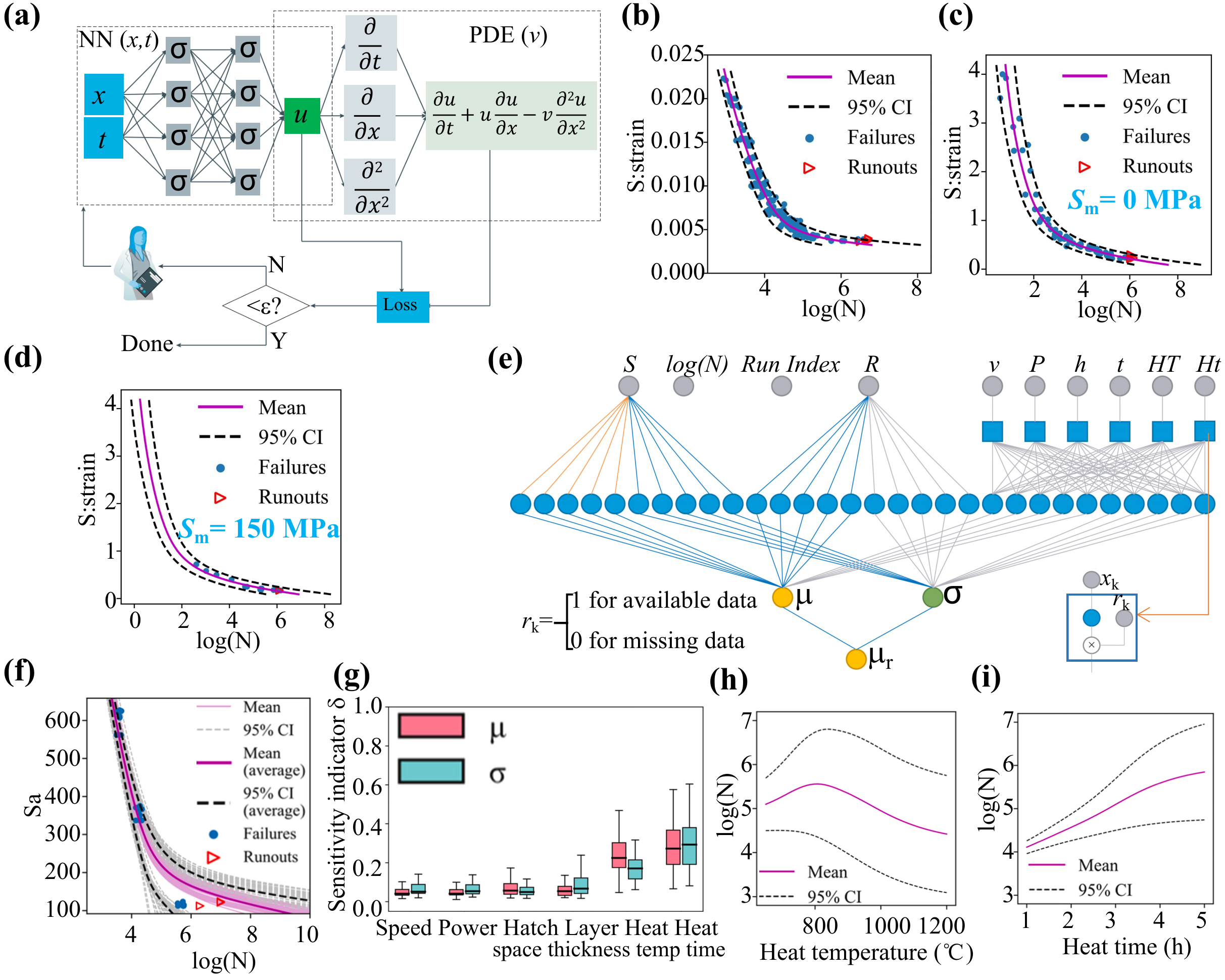}
	\caption{(a) Illustration of PINN for solving viscous Burger's equation~\cite{Karniadakis2021}; (b) Validation of PPgNN using single-factor experimental data ($S$) for nickel-base superalloy~\cite{CHEN2021114316}; (c) (d) Validation of PPgNN using multi-factor experimental data ($S$ and $S_m$) with the consideration of mean stress effect~\cite{CHEN2021114316}; (e) PPgNN for fatigue analysis of SLM-AM Ti64 with the consideration of missing data~\cite{CHEN2021101876}; (f) PPgNN predicted 50 S--N curves and their average curve~\cite{CHEN2021101876}; (g) Sensitivity indicator for processing and post-processing parameters~\cite{CHEN2021101876}; (h) (i) Parametric study on the influence of post-processing parameters on predictd probabilistic fatigue lives~\cite{CHEN2021101876}. }
	\label{fig07}
\end{figure*}

In contrast to the classic NN models that are purely data-driven, the physics-informed neural network (PINN) takes into account the underlying physics of the problem. PINN produces results that possess well-known physics knowledge~\cite{BELFIORE20071705}.
A typical application of PINN is to solve mathematics problems such as partial differential equations (PDEs)~\cite{cuomo2022scientific}, e.g., viscous Burger's equation.
As shown in Fig.~\ref{fig07}(a), the left part is physics-uninformed and represents the surrogate of the PDE solution $u(x,t)$. The right part is physics-informed and describes the PDE residual.
The loss function $L$ includes a supervised loss of data $u$ from the initial and boundary conditions ($L_{data}$) and an unsupervised loss of PDE ($L_{PDE}$)~\cite{Karniadakis2021}. Taking viscous Burger's equation as an example in Fig.~\ref{fig07}(a), one has
\begin{gather}
	L=w_{data}L_{data}+w_{PDE}L_{PDE} \\
	L_{data}=\frac{1}{N_{data}}\sum_{i=1}^{N_{data}}\left(u\left(x_i,t_i\right)-u_i\right)^2 \\
	L_{PDE}=\frac{1}{N_{PDE}}\sum_{j=1}^{N_{PDE}}\left(
	\frac{\partial u}{\partial t}+u\frac{\partial u}{\partial x}-v\frac{\partial^2 u}{\partial x^2}\right)^2\vert_{(x_j,t_j)},
\end{gather}
in which $(x_i,t_i)$ and $(x_j,t_j)$ are two sets of points for the initial/boundary conditions and for the entire domain, respectively. $u_i$ are values of $u$ at $(x_i,t_i)$. $w_{data}$ and $w_{PDE}$ are the weights for balancing the interplay between the two loss terms ($L_{data}$ and $L_{PDE}$). These weights could be tuned automatically or defined by user, and play a critical role in improving the trainability of PINNs~\cite{202211}. The PINN is trained by minimizing the loss until the loss is smaller than a specified threshold.

A direct application of black-box NN models to fatigue life prediction could suffer from some challenges, since the statistical relations between inputs and outputs that are solely learned from the data could possibly violate some commonly known physics laws or knowledge. Additionally, for scenarios outside the training data range, overfitting could occur in the NN models, leading to unreasonable results. In order to overcome the above challenges, Chen et al.~\cite{CHEN2021114316} propose a probabilistic physics-guided neural network (PPgNN) for the estimation of probabilistic fatigue S--N curve. The input layer is compromised of logarithmic of fatigue life $\log(N)$, stress level $S$ or strain level $\varepsilon$, and an index to indicate whether a data is a failure or runout. Unlike the regular NN that only learns the mean from the collection of distributed data, the output layer in PINN has two output neurons, including both the mean $\mu$ and the standard deviation $\sigma$.

In order to make sure that the NN could produce physically reasonable results, there are two main considerations on the knowledge in predicting the fatigue life with an applied stress or strain. The first knowledge is that the desired S--N curves should present the increased scatter with the decrease of stress level, i.e., $\sigma$ should increase as the stress or strain decreases. The second knowledge is that the S--N curves should show an infinite life or sufficiently large number of cycles at low stress levels, i.e., the curvature of the fatigue curve should decrease as the stress decreases and there exists an asymptotic behavior near the fatigue limit or near very long life if an apparent fatigue limit does not present~\cite{doi:10.1080/00401706.1999.10485925}. 
Fot the first knowledge, one requires 
\begin{equation}
	\frac{\partial\sigma}{\partial S}\leq 0
	\label{kno1}
\end{equation}
and the second knowledge could be satisfied if
\begin{equation}
	\frac{\partial^2\mu}{\partial S^2}\geq 0.
	\label{kno2}
\end{equation}
Those equations can be transformed to put constraints on weights and biases of NN, thus resulting in the so-called PINN. Due to the over restrictive constraints on the weight and bias, in the PINN architecture another layer with a relaxed mean ($\mu_r$) is added. Then the final outputs are $\mu_r$ and $\sigma$.

The probabilistic S--N curves for nickel-base superalloy are predicted by PPgNN, as shown in Fig.~\ref{fig07}(b)~\cite{CHEN2021114316}. The predicted S--N curves present satisfactory physics and knowledge: the curvature of the fatigue curve decreases and the standard deviation of fatigue life increases as the strain decreases, and there appears an asymptotic behavior near the fatigue limit. Moreover, it is convenient to incorporate other influencing factors into the PPgNN. As shown in Fig.~\ref{fig07}(c) and (d), the effect of mean stress ($S_m$) by setting $S_m=0$ and 150 MPa is considered. It can be seen that for each case, almost all testing data points are located within the 95\% confidence interval, and the above two types of physics and knowledge can be satisfied.

Furthermore, Chen et al.~\cite{CHEN2021101876} consider the effect of AM processing parameters (scanning velocity $v$, laser power $P$, hatch distance $h$, powder layer thickness $t$, heating temperature $H_T$, and heating time $H_t$) on fatigue properties of SLM-AM Ti64 through PPgNN. In accordance with the previous work~\cite{CHEN2021114316}, an inclusion of physics knowledge (Eq. \ref{kno1} and Eq. \ref{kno2}) is realized by imposing constrains on weights and biases between fatigue parameters (stress amplitude $S$ and stress ration $R$) inputs and outputs. There is no constraint imposed on biases and weights between AM processing parameter inputs and outputs, since the physics knowledge between AM processing parameters and fatigue life (mean $\mu$ and standard deviation $\sigma$) is unclear and this knowledge is intended to be obtained from PPgNN.
In their work, more than half of the datasets contain missing data, i.e., 6 datasets are complete and 12 datasets contain missing data about the SLM-AM processing parameters. Efforts are required to make full use of these incomplete datasets rather than brutally abandon such datasets that could contain potentially valuable information. As shown in Fig.~\ref{fig07}(e), the squarely shaped neurons named as selective neurons are designed to make NN trained with the incomplete processing parameter datasets. If the data is available ($r_k=1$), the selective neuron will be active in the computation of the network outputs. If the data is missing ($r_k=0$), the selective neuron would work with the neglection of input data~\cite{Lopes2012357}.

As shown in Fig.~\ref{fig07}(f), the predicted S--N curves are obtained by using 50 sets of initial random values for bias and weight before training the model. Due to the randomness of the initial values, the results generated by NN are not constant. The average of calculated 50 S--N curves is plotted as a thick line (mean S--N curve) in Fig.~\ref{fig07}(f). All the experimental data are found to locate within the predicted confidence bounds. In order to obtain more reliable results, it is necessary to train the NN for multiple times. The influence of processing parameters on both the mean and standard of the fatigue life under the fixed fatigue parameters (350 MPa for stress amplitude and 0.1 for stress ratio) is summarized in Fig.~\ref{fig07}(g). Within the range of collected data, the mean fatigue life increases with the laser scanning speed and layer thickness, but decreases with the increase of laser power and hatch space.
The fatigue life is found more sensitive to the post-processing parameters such as heat temperature and heat time than to the SLM-AM in-processing parameters. Fig.~\ref{fig07}(h) and (i) summarize the effect of heat time and heat temperature on the probabilistic fatigue lives. The mean fatigue life increases with the heat time, but increases and then decreases with the increase of heat temperature.

\section{Support vector machine for fatigue life of AM materials}
Support vector machine (SVM) is a supervised ML algorithm that relies on the principle of structural risk minimization~\cite{Noble2006,Cortes1995}. The SVM can map the inputs to high-dimensional feature space (Hilbert space) by using the kernel function, and thus could result in high efficiency in regression analysis and nonlinear classification. When applied to regression, the method is also called the support vector regression (SVR)~\cite{SANCHEZA20035}.
For example, an input vector $x=\left\{ x_1,...,x_n \right\}$ and an output vector $y=\left\{ y_1,...,y_n \right\}$ are used in regression analysis, and one has the expression
\begin{equation}
	f(x)=w\varphi(x)+b,
\end{equation}
where $f(x)$ is a regression function, and $w$ and $b$ are the weight coefficient and bias, respectively. $\varphi(x)$ is the function that maps the variables from the original space to a high-dimensional feature space and thus enables linear regression of samples in this feature space. The optimization aim of SVR can be expressed as~\cite{DANG2022106748}
\begin{equation}
	\text{min}\left\{\frac{1}{2}w^2+C\sum_{i=1}^{n}\xi_i+\xi_i^*\right\},
\end{equation}
where $C$ is a penalty parameter, and $\xi_i$ and $\xi_i^*$ are relaxation factors. The constraints imposed to optimization process are given by
\begin{equation}
\begin{cases}
	y_i - w \varphi(x_i)-b \leq \epsilon + \xi_i   \\
	y_i - w \varphi(x_i)-b \geq \epsilon + \xi_i^* \\
	\xi_i \geq 0, \xi_i^* \geq 0, i=1,2,...n
\end{cases},
\end{equation}
where $\epsilon$ represents the tolerable deviation between the predicted and actual value. To optimize parameters, some methods such as grid search method and Bayes search method can be used~\cite{LUO2021140693}.

SVM is also tried to predict fatigue properties of AM materials. For predicting the fatigue life of Ti6.5Al2ZrMoV titanium manufactured by laser DED (direct energy deposition), Dang et al.~\cite{DANG2022106748} utilize a SVR algorithm by post-mortem fractography analysis.
Because the fatigue life of AM materials is mainly influenced by their microstructures and defects, they focus on the fatigue life of AM metals with microstructure pores. The pores are divided into four types according to the crack initiation modes. Small pores possess a diameter less than the diameter of the small-size $\alpha$ phase, typically 20 $\mu$m. Medium pores possess a diameter greater than the diameter of the small-size $\alpha$ phase and less than the diameter of a micro-columnar grain, typically 20-60 $\mu$m. Large pores with easily observed facets possess a diameter greater than the diameter of a micro-columnar grain, and the facets constitute most of the fracture surface near the pore. Large pores with less identifiable facets are not easily observed, and the facets constitute a small part of fracture surface near the pore.
The range of stress intensity ($\Delta K$) is used to evaluate the stress around the pores in Murakami's approach. In addition, ratio of the distance to a free surface to the pore equivalent diameter ($r_d$) is used to define the location of a pore. The pore size ($area$) and the experimental peak stress ($\sigma_{\text{max}}$) are selected as the input variables for SVM. The correlation coefficient $r$ and mean percentage error (MPE) are utilized to estimate the SVM accuracy.
SVM accuracy with different input variables is summarized in Table \ref{table1}, in which the pores are only divided into small, medium, and large pores for SVR6 model.
It is found from SVR1 model and SVR2 model that the accuracy of the model would be low if there is no failure-mode analysis of material in the pores' vicinity. The accuracy of SVR3 model is also low when $\Delta K$ is not used as input parameter. These results imply the significance of extracting pore types and input variables $\Delta K$. Moreover, the SVR6 model possesses an increased MPE and a decreased $r$, indicating the necessity of considering the cleavage around large pores in the fatigue life prediction.

\begin{table}[!t]
	\caption{\label{table1} Errors of SVR models with different input variables for predicting fatigue life of Ti6.5Al2ZrMoV titanium manufactured by laser DED~\cite{DANG2022106748}.}
	\centering
	{
		\begin{tabular}{ccccc}
			\toprule
			&Input variables& Test MPE&Train MPE&$r$\\
			\midrule
			SVR1&$\Delta K$, pore types&0.3117&0.2117&0.8222\\
			SVR2&$\Delta K$&0.4596&0.3077&0.6916\\
			SVR3&$\sigma_{\text{max}},area,r_d$, pore types&0.3629&0.3360&0.6419\\
			SVR4&$\sigma_{\text{max}},area$, pore types&0.3828&0.3738&0.6370\\
			SVR5&$\Delta K,r_d$, pore types&0.3710&0.3511&0.6465\\
			SVR6&$\Delta K$, pore types$^*$&0.3728&0.2770&0.6402\\
			\bottomrule
	\end{tabular}}
\end{table}

\begin{figure*}[!b]
	\centering
	\includegraphics[width=16cm]{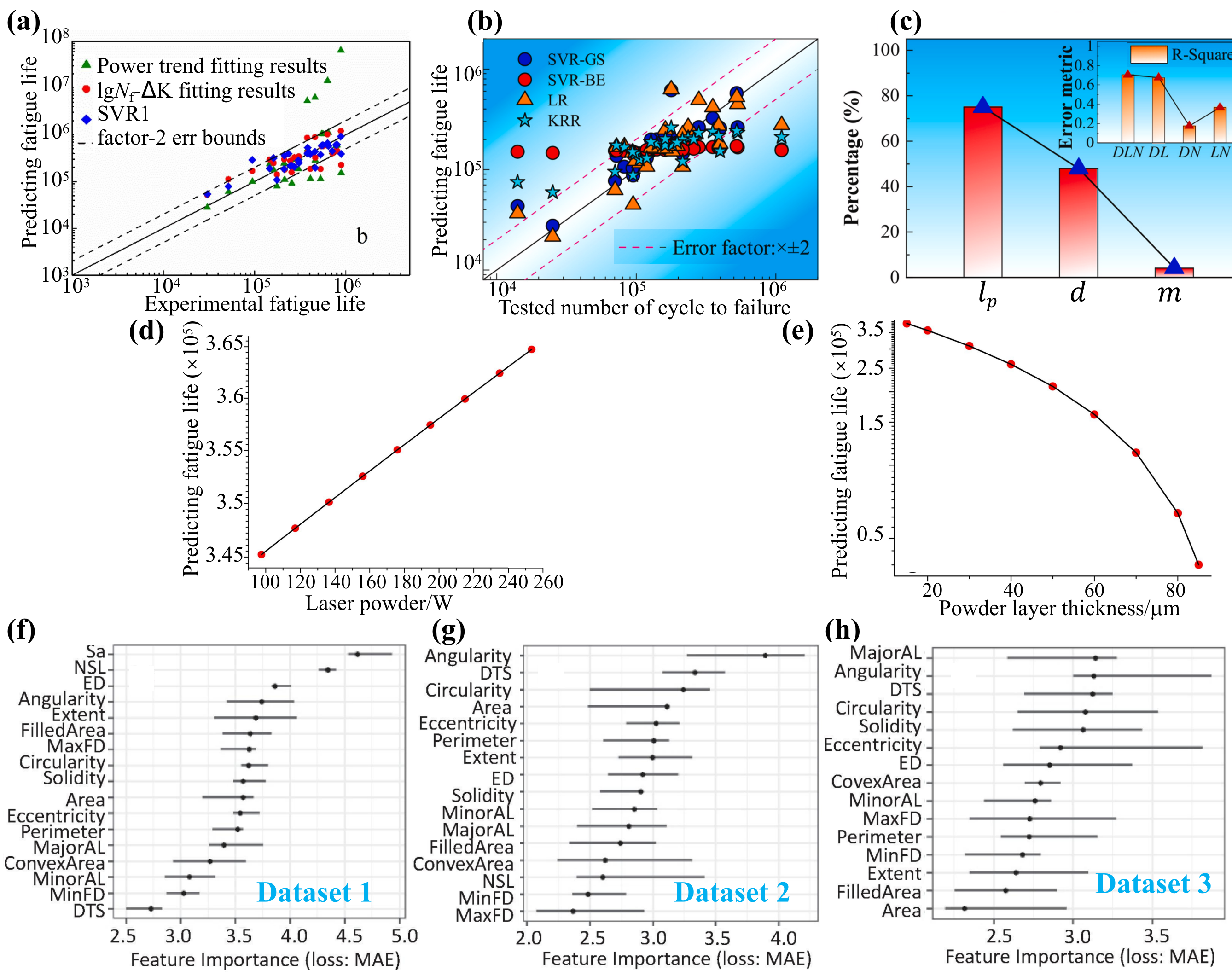}
	\caption{(a) Experimental and predicted fatigue life by SVR and fitting results for Ti6.5Al2ZrMoV titanium manufactured by laser-DED~\cite{DANG2022106748}; (b) Fatigue life from predictions by different ML algorithms and experimental tests for SLM-AM Inconel 718~\cite{LUO2021140693}; (c) Influence of different kinds of pore features on the SVR-GS algorithm performance~\cite{LUO2021140693}; Predicted fatigue life by SVM models for AM 316L stainless steel with (d) different laser power and (e) different powder layer thickness~\cite{ZHAN2021105941}; (f)-(h) Permutation feature importance of defect features based on SVR for LPBF-AM 17-4 PH stainless steel specimens by using different dataset (SA: Strain amplitude; NSL: Normalized stress level; DTS: Distance to the surface; ED: Equivalent diameter: FD: Feret diameter; AL: Axis length)~\cite{LI2022107018}.}
	\label{fig09}
\end{figure*}

The SVM results for predicted fatigue life could be compared to two regression equations, i.e., the linear relationship~\cite{WANG2019373}
\begin{equation}
	\lg N_f=a+b\Delta K
	\label{eqlgNf}
\end{equation}  
and the power relationship~\cite{SHERIDAN2021106033}
\begin{equation}
\Delta K=A\cdot\left(N_f\right)^m+c.
\label{eqDeltaK}
\end{equation}
Fig.~\ref{fig09}(a) presents the comparison between experimental and predicted fatigue life from the SVR1 model and fitting results by Eqs. (\ref{eqlgNf}) and (\ref{eqDeltaK}). The errors of the power-trend regression are far greater than those of SVR1 model and the linear $\lg N_f-\Delta K$ regression. In SVR1 model there exists only one predicated point beyond the two-error bounds, suggesting that SVR1 model should have better accuracy than regression equations in fatigue life prediction.

In addition, Luo et al.~\cite{LUO2021140693} utilize SVR algorithms to predict the fatigue life of SLM-AM Inconel 718. After observing and analysing the pores on the fractured surfaces of specimens, they select three types of pore features as input variables: the number of pores ($m$), the maximum pore diameter ($d$), and the pore location (nearest neighbored distance to quantitatively describe the pore location, represented by $l_p$). Besides, the stress amplitude ($\sigma$) of bending fatigue is also taken as an input variable. To optimize the SVR parameters, two different methods including gird search (SVR-GS) and Bayes search (SVR-BS) are utilized.
Fig.~\ref{fig09}(b) presents the comparison between the predicted fatigue life obtained by different algorithms (SVR-GS, SVR-BS, linear regression (LR), and kernel ridge regression (KRR)) and the experimentally determined fatigue life. Among all algorithms, the SVR-GS algorithm and LR algorithm present superior performance than others, but the LR algorithm does not work well in the 'lack of data' regime. SVR-GS algorithm is found suitable for the prediction of both low cycle and high cycle fatigue life.
In Fig.~\ref{fig09}(c), the influence of each individual pore feature and different combinations of pore features is quantitatively examined. The inset of Fig.~\ref{fig09}(c) presents the variation of $R^2$ when the model takes different types of pore features, in which $D$, $L$, and $N$ represents pore diameter, pore location, and pore number, respectively. It can be found from Fig.~\ref{fig09}(c) that the pore location has the most prominent influence on the fatigue life, while the pore number has a little impact on the fatigue life.
Similarly, Bao et al.~\cite{BAO2021107508} also utilize SVM to explore the influence of defect location, size, and morphology on the fatigue life of a SLM-AM Ti6Al4V alloy. They find that in comparison with defect size and morphology, the defect location has much stronger effect on the fatigue life, agreeing with the conclusion from Luo et al.~\cite{LUO2021140693}.

In order to consider the effect of processing parameters on fatigue life of AM 316L stainless steel, Zhan et al.~\cite{ZHAN2021105941} develop a SVM-based model for fatigue life prediction. As shown in Fig.~\ref{fig09}(d), the laser power increasing from 100 to 200 W only slightly increases the fatigue life, implying that the effect of laser power on the enhancement of fatigue life of AM 316L stainless steel is not apparent. However, the increase in the powder layer thickness ($t$) from 20 to 80 $\mu$m could significantly decrease the fatigue life from 3,000,000 to 30,000, as shown in Fig.~\ref{fig09}(e). These results indicate that for improving the fatigue performance, decreasing the powder layer thickness is more efficient than modifying the laser power.

Furthermore, in order to explore the relationship between the geometrical features of critical defects measured from fracture surface and the fatigue performance of LPBF-AM components, Li et al.~\cite{LI2022107018} develop a kernel SVR model.
In detail, they categorize the defect features into morphology-, size-, distance-, and process-related features. All these features are utilized to quantify various aspects of defects and to investigate the connection between defects and fatigue life of LPBF-AM 17-4 PH stainless steel specimens. Because the prediction accuracy of SVM relies on both the data quality and data size, they select three irrelevant datasets for analysis. Dataset 1 with large data size but low consistency includes 157 defects on all specimens. Dataset 2 with medium data size and medium consistency includes 51 defects on specimens that have different heat treatments but same strain amplitude. Dataset 3 with small data size but high consistency includes 34 defects on specimens that have the same heat treatment and same strain amplitude.
Permutation feature importance (PFI) is a suitable indicator to evaluate the importance of a feature by calculating the increase in the SVR model error after permuting the feature. As shown in Fig.~\ref{fig09}(f), the normalized stress level (NSL) and strain amplitude (SA) parameters are predominant features in Dataset 1; because Dataset 1 includes all specimens that have different heat treatments and strain amplitudes. For Dataset 2 in Fig.~\ref{fig09}(g) and Dataset 3 in Fig.~\ref{fig09}(h), even though large variations exist in the importance of defect features, they are not statistically significant.

Then the impacts of defect features on fatigue life of specimens for Dataset 2 and Dataset 3 are further explored~\cite{LI2022107018}. For Dataset 2, increasing all the size-related features (e.g., major axis length, area, max feret diameter, and perimeter) could decrease the fatigue life. As for morphology-related defect features, when the distance to the surface and the size-related features are the same, the fatigue life of specimens tends to be longer if the critical defects have a larger radius (eccentricity), are more circular (circularity), and is more convex with a smoother contour (solidity). It is also revealed that the critical defects that are far away from the surface would result in longer fatigue life. In Dataset 3, except for that no clear trend appears for the relationships between perimeter and fatigue live, other features still notably influence the fatigue life of LPBF-AM specimens like in Dataset 2.

\begin{figure*}[!b]
	\centering
	\includegraphics[width=16cm]{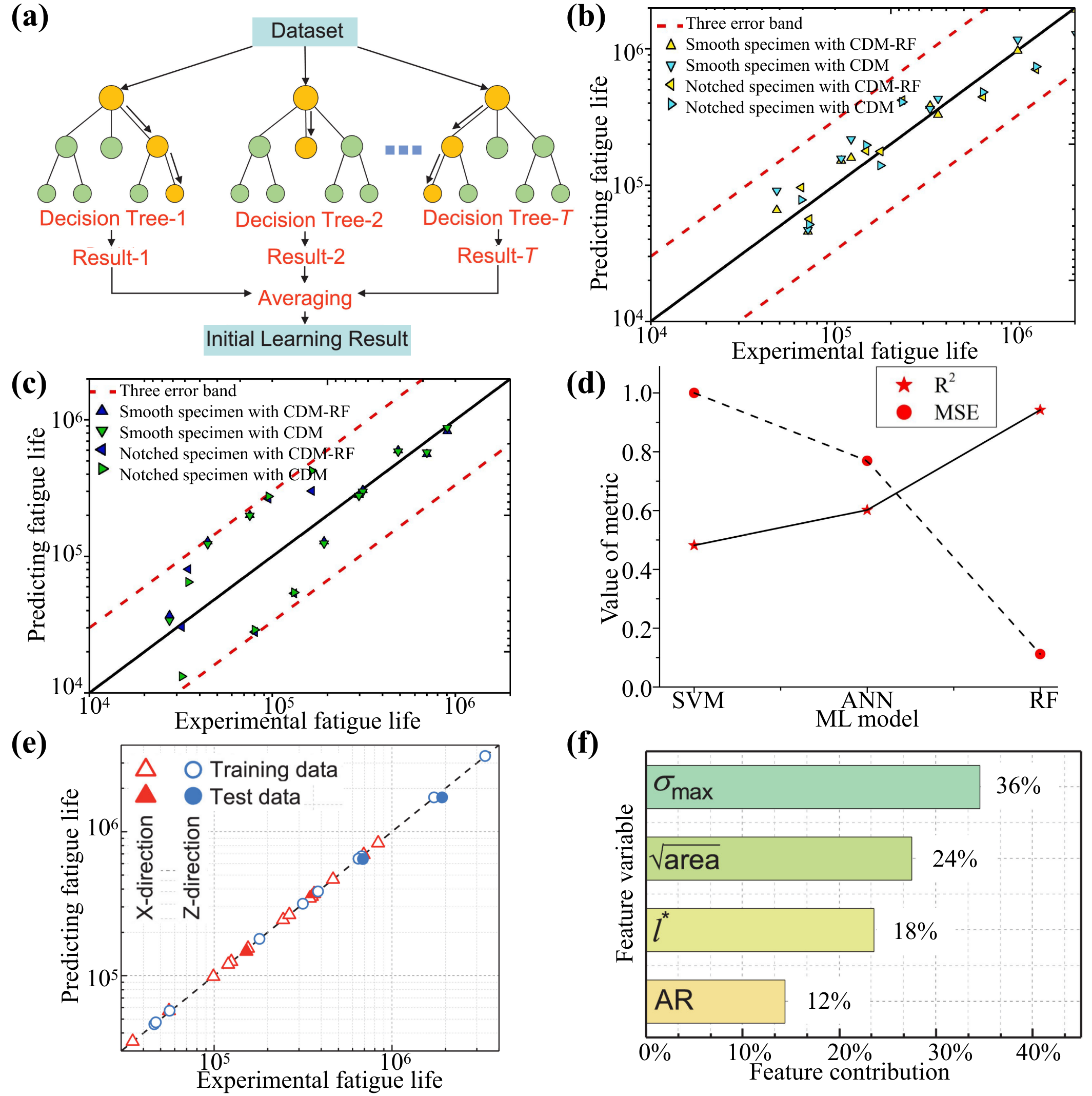}
	\caption{(a) Schematics of a RF model structure~\cite{PENG2022107185}; Fatigue life predicted by CDM-RF and CDM for AM (b) TA2, TA15 and (c) TC4, TC11~\cite{ZHAN2021107850}; (d) Prediction performances of different ML models (SVM, ANN, RF) for 316L AM stainless steel~\cite{ZHAN2021105941}; (e) RF model prediction of fatigue life using the training data and test data for LPBF-AM AlSi10Mg samples that are tested along (Z-direction) and perpendicular to (X-direction) the build direction using RF model~\cite{PENG2022107185}; (f) Percentage contribution of the top four features to the predicted fatigue life of LPBF-AM AlSi10Mg by RF model~\cite{PENG2022107185}.}
	\label{fig11}
\end{figure*}

\section{Random forest for fatigue life and fatigue crack growth rate of AM materials}
The random forest (RF) algorithm is firstly proposed by Breiman to deal with the classification and regression problems~\cite{Breiman2001}.
As illustrated in Fig.~\ref{fig11}(a), a number of decision trees have to be created in the RF model, thus forming an upside-down tree. The root and leaves correspond to the top of the tree and the terminal nodes, respectively. So a decision tree could be treated as a binary tree in which each intermediate node has two outputs. Furthermore, a decision tree usually consists of a number of nodes to evaluate the inputs by the test function. Then the values are transmitted to the tree branches with regard to the sample characteristics. Multiple uncorrelated decision trees are constructed during the training stage of RF regression model. The final predicted results are obtained by averaging the outputs of all decision trees~\cite{ZHAN2021107850}.

For the data-driven prediction of fatigue life of AM titanium alloys such as TA2, TA15, TC4, TC11 etc., Zhan et al.~\cite{ZHAN2021107850} present a CDM based RF model. The bootstrap aggregating algorithm is used to train the RF regression model. Three input variables (maximum stress $\sigma_{\text{max}}$, stress concentration factor $K_t$, and stress ratio $R$) and one output variable (fatigue life $N_f$) are involved.
The influence of RF parameters (the total number of trees $n_t$, the maximum depth of the tree $n_d$, and the number of variables $n_v$ in each tree) on the prediction accuracy is examined. It is found that the values of $n_t \geq 20$ and $20 \leq n_d \leq 40$ are beneficial for a better prediction accuracy and efficiency. In contrast, the effect of $n_v$ on prediction performance is not significant. Furthermore, the results predicted by the CDM-RF method are compared with that by the CDM method, as shown in Fig.~\ref{fig11}(b) and (c). It can be seen that for AM titanium alloys, the DM-RF method exhibits better prediction performance than the CDM method. For the AM TA2 and TA15, the results predicted by the CDM-RF method are much closer to the experimental data than those by the CDM method, especially for the notched specimens in Fig.~\ref{fig11}(b). In terms of the AM TC4 and TC11 in Fig.~\ref{fig11}(c), the improvement in the prediction performance by CDM-RF method is not obvious for both the smooth and notched specimens. In general, the predicted performance for AM TC4 and TC11 (R$^2$=0.875) is better than that of the AM TA2 and TA15 (R$^2$=0.682).
Moreover, the performances of various ML models (e.g. FNN, SVM, RF, etc.) for predicting the fatigue life of AM 316L stainless steel parts is alsp examined~\cite{ZHAN2021105941}. As shown in Fig.~\ref{fig11}(d), the prediction performance of the FNN model are better than that of the SVM model. In comparison with the FNN and SVM models, the RF model shows the highest accuracy and best performance in the fatigue life prediction of AM 316L stainless steel.

\begin{figure*}[!b]
	\centering
	\includegraphics[width=16cm]{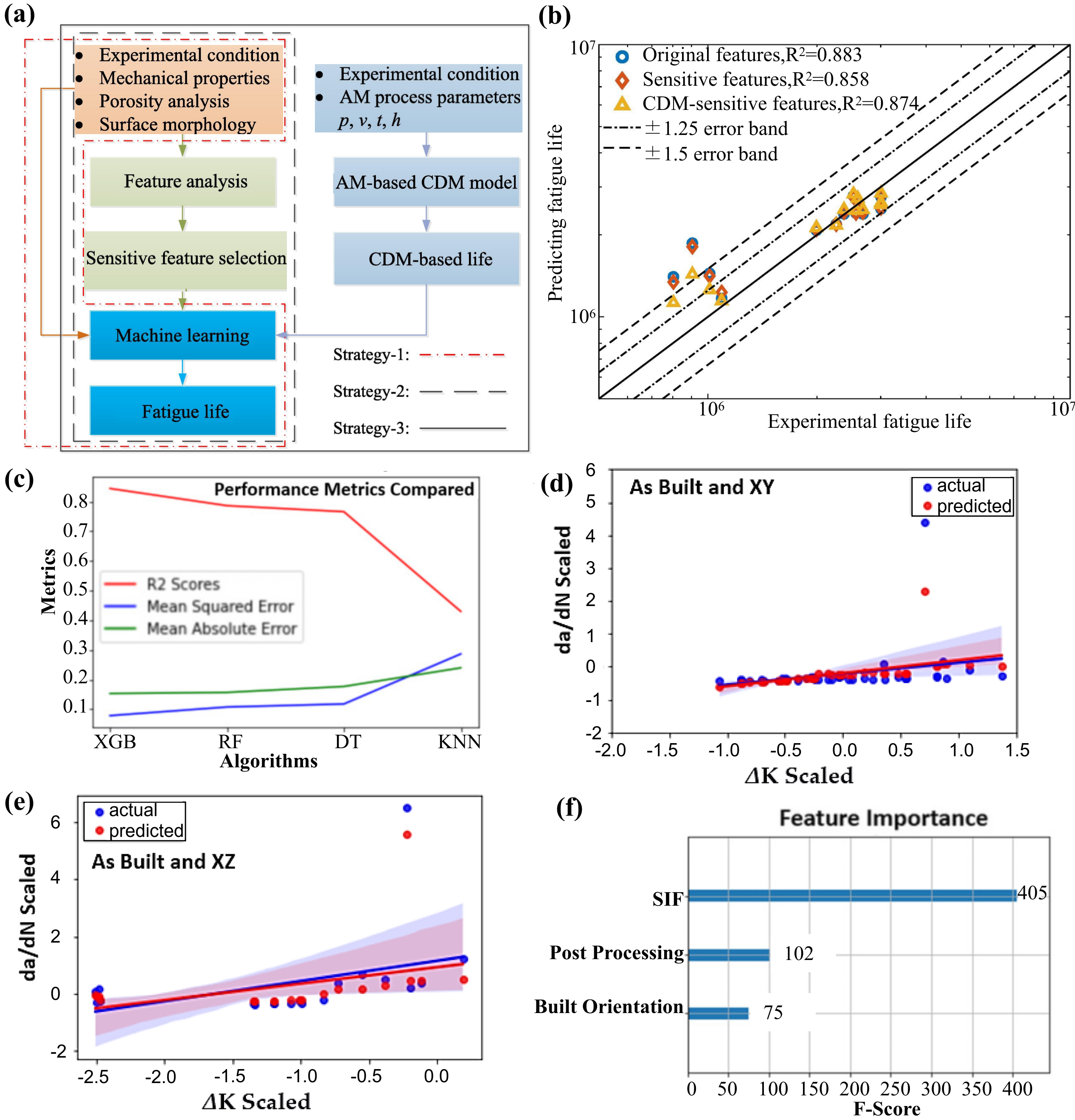}
	\caption{(a) Strategy diagram of RF-based fatigue life prediction model~\cite{WANG2022107147}; (b) Fatigue life prediction by RF model using three strategies for LPBF-AM AlSi10Mg~\cite{WANG2022107147}; (c) Performance metrics of four different ML models for predicting fatigue crack growth rate of LPBF-AM Ti6Al4V~\cite{met12010050}; Experimental and predicted $\text{d}a/\text{d}N$ curves for as built AM Ti64 alloy with (d) XY and (e) XZ specimen orientations~\cite{met12010050}. (f) Feature importance ranking among stress intensity factor (SIF), post processing, and built orientation~\cite{met12010050}.}
	\label{fig12}
\end{figure*}

For predicting the fatigue life of LPBF-AM AlSi10Mg alloy, Peng et al.~\cite{PENG2022107185} develop a RF based approach. For each fatigue sample, features including size, morphology, and location of the critical defect at the crack initiation sites are quantified from the post mortem fractography to form a dataset. Considering features that are potentially important in determining the fatigue life, five types of geometrical features from the critical defects are considered: Murakami's defect size parameter that is defined by the defect area projected onto the crack plane ($\sqrt{area}$), the maximum stress ($\sigma_{\text{max}}$), the aspect ratio of the ellipse that approximates the defect (AR), the angle between the major axis and the sample surface tangent ($\Phi$), the defect jaggedness ($J$), and the distance of defect location to the surface ($l^*$).
Since the sample size is relatively small, applying all the defect features to train the RF model would lead to a weak generalization. Therefore, the top four most important features from the RF model possessing six features (R$^2$=0.78) are chosen, i.e., $\sigma_{\text{max}}$, $\sqrt{area}$, $l^*$, and AR, as input feature variables to optimize the RF model (R$^2$=0.85).
As shown in Fig.~\ref{fig11}(e), the optimized RF model can predict the fatigue life of the test-pieces based on the applied stress, projected area of critical defect, defect aspect ratio, and defect location. The prediction is shown to possess good generalizability and high accuracy.
The optimized RF model also readily predict the fatigue life of samples tested parallel and perpendicular to the build direction.
The much lower fatigue life for the tests along the build direction than those perpendicular to it could be attributed to the different projected shapes ($\sqrt{area}$ and AR) of the crack-like defects in the two cases.
The contribution of four features to the high cycle fatigue life of LPBF-AM AlSi10Mg alloy is summarized in Fig.~\ref{fig11}(f). It can be seen that $\sigma_{\text{max}}$ and $\sqrt{area}$ are the dominant features in determining fatigue life, and the effect of four parameters on fatigue life decreases in the order: $\sigma_{\text{max}}>\sqrt{area}>l^*>\text{AR}$. It suggest that for the same external loading, decreasing the defect size or Murakami's area in AM materials is critical for improving the fatigue performance.

To examine the effect of causality among above features on the prediction accuracy of the RF model, Wang et al.~\cite{WANG2022107147} combine CDM theory with the effect of AM parameters to construct the RF model by three strategies for predicting the fatigue life of LPBF-AM AlSi10Mg alloy. As shown in Fig.~\ref{fig12}(a), in Strategy-1 the original features are directly used as inputs of RF model to construct the fatigue life prediction model. In Strategy-2, the sensitive features are extracted from the 12 original features (stress amplitude, ultimate strength, yield strength, surface hardness, elongation, depth of hardness, surface compressive residual stresses, depth of compressive residual stresses, maximum compressive residual stresses, relative density, surface modification factor, and surface roughness) to reduce the interdependence among features. Strategy-3 is based on Strategy-2, but the predicted results from AM-based CDM are used as physical information to characterize the damage process of AM parts.
The weights of these 12 features are calculated by the gradient boosting decision tree (GBDT) algorithm. It is found that the sensitive features of the RF-based life prediction model are surface roughness, stress amplitude, surface compressive residual stresses, and maximum compressive residual stresses. Fig.~\ref{fig12}(b) presents the predicted results by RF model using those three different strategies. The Strategy-3 based on CDM-sensitive features is found to possess the highest prediction accuracy. The Strategy-2 based on sensitive features exhibits better prediction performance than those based on original features (Strategy-1). It is suggested that RF model guided by physical knowledge and sensitive features could have better generalization performance and prediction accuracy in fatigue life prediction of AM materials~\cite{WANG2022107147}. 

For predicting the fatigue crack growth rate ($\text{d}a/\text{d}N$) of LPBF-AM Ti6Al4V with post processing, Konda et al.~\cite{met12010050} have tried four ML algorithms including RF, decision trees, extreme gradient boosting, and $k$-nearest neighbor. Based on the domain knowledge, the post-processing technique (as-built, stress relief, and heat treatment), the stress intensity factor range ($\Delta K$), and build orientation (XY, XZ, and ZX) are selected as the input features.
As shown in Fig.~\ref{fig12}(c), the performance metrics of all these four algorithms are compared. It can be seen that RF and extreme gradient boosting method perform better than decision tree and $k$-nearest neighbor methods. Specifically, in terms of least mean squared errors (MSE) and higher R$^2$ scores, extreme gradient boosting method is superior to other methods for the prediction of fatigue crack growth rate. The predicted $\text{d}a/\text{d}N$ in Fig.~\ref{fig12}(d) and (e) by extreme gradient boosting method is found to agrees well with the experimental data. Furthermore, the input features importance analysis in Fig.~\ref{fig12}(f) indicate that the most influencing feature is the stress intensity factor range, and the post-processing technique is more important than the built orientation.

\section{Summary and outlooks}
In summary, we have presented a comprehensive overview on the latest progress of predicting fatigue properties (fatigue life and fatigue crack growth rate) of AM materials by various ML strategies.
ML strategies including FNN, CNN, ANFS, SVM, and RF have been applied to predict the fatigue life of AM metals (Ti alloy, Al alloy, stainless steel, Ni-based superalloy, etc.), as well as their correlations with microstructures, AM processing parameters, and post-processing parameters.
In particular, RF has been used to predict the fatigue crack growth rate of AM Ti64 alloy.
Nonetheless, the application of ML to predict fatigue properties of AM materials is still rare. A collection of fatigue data of AM materials treated with various post-processing techniques from published literatures is meaningful. All these ML strategies are demonstrated to achieve satisfactory results in the prediction of fatigue properties of AM materials. In addition, different ML strategies could achieve similar prediction performance for a specific mechanical property of a specific AM material if they are reasonably trained and designed. There currently seems no unified ML strategy for all cases.
According to the state-of-the-art reports in the literatures, ML has undoubtedly emerged as a powerful tool for the efficient and accurate prediction of fatigue properties of AM materials.
Nevertheless, there exist the following challenges and the associated potential solutions.

\textbf{\textit{Small dataset} of fatigue properties of AM materials.}
Accessible dataset is directly related to the performance of ML algorithms in predicting fatigue properties of AM materials. An obvious drawback of ML model is that training the model requires a large amount of data. However, it is always expensive and time consuming to collect a large number of experimental datasets~\cite{WOS:000518708500028,doi:10.2514/6.2020-0680}. For AM, this case is even serious, since the data on fatigue properties of AM materials is currently limited or even unaccessible for some critical materials. Especially, there are only a few reports on the data of fatigue crack growth rate. Generating experimental data for the fatigue properties of AM materials by considering different material types, AM techniques, AM processing parameters, post-processing parameters and microstructural features is extremely high-cost and time consuming.
Many methods have been proposed to deal with small dataset problems. For example, data augmentation as a representative technology could generate new dataset by adding a small disturbance to the original dataset~\cite{WOS:000604573000113,WOS:000074842400012}. Many generative models, such as variational autoencoder~\cite{kingma2013autoencoding} and adversarial autoencoder~\cite{makhzani2015adversarial}, can provide ways to perform data augmentation. Another representative method to generate dataset is based on physics models. For instance, fatigue data from finite-element simulation~\cite{Tang2023Modeling,Min2021Computational} with virtual microstructures of AM materials can be used to expand the training dataset. Moreover, there is little data on the very high cycle fatigue of AM materials and on the fatigue crack behavior around the threshold value.

\textbf{\textit{Multifarious features} in ML models for predicting fatigue properties of AM materials.}
AM is a very complex process, which contains multifarious features including processing and post-processing parameters, as well as numerous and diverse microstructure characteristics. Some of these features could have significant effects on the fatigue properties of AM parts, while others may have little effect. 
Therefore, it is crucial to select a good set of features for ML models to achieve good prediction performance~\cite{1581138385}. On one hand, the principles of feature selection depend on the researchers' experience and knowledge on AM techniques and AM materials. There is no unified rule to determine which feature is better. For example, some important processing parameters including laser power, layer thickness, and hatch distance can be selected as ML input parameters. On the other hand, some ML algorithms such as RF and SVR can evaluate the importance of different individual features and their combination, providing a guidance for selection from multifarious features with little experience or knowledge about AM and its induced microstructure.

\textbf{\textit{Overfitting} for scenarios outside the training fatigue data range of AM materials.}
A ML algorithm has advantages in the generalization ability for predicting fatigue properties of AM materials with various AM material types, AM techniques, AM processing parameters, post-processing parameters, microstructural features, etc.
If the model provides good prediction performance for the training data but shows poor prediction performance for the validation dataset, overfitting happens and unreasonable fatigue life will be predicted. To avoid overfitting, some techniques including regularization~\cite{e23121673,8402232}, early stopping~\cite{PRECHELT1998761,song2019does}, and dropout~\cite{XieSEPT1} are developed. For instance, if MSE is a function of training process, MSE of the training dataset decreases during the whole training process. MSE of the validation dataset first decreases and then increases. Thus, the training process should early stop when MSE of the validation dataset reaches the minimum value in order to avoid overfitting of the data on fatigue properties of AM materials.

\textbf{\textit{Low interpretability} of ML models for predicting fatigue properties of AM materials.}
Classic ML models are purely data-driven methods, in which the input-output relations cannot be directly attained due to the black box nature.
How to interpret the predicted fatigue properties of AM materials still remains as a issue. The output fatigue life from the ML models is difficult to correlated with the underlying fatigue mechanism in AM materials. As mentioned above, PINN is a new approach for combing scientific knowledge with data science to improve the interpretability of ML models. Some physics knowledge can be designed as the loss function or parameters of ML models~\cite{QI2019721}. How to add more physical knowledge and fatigue mechanisms about AM techniques, AM materials, mechanics models, microstructure and multiscale features into the PINN model for an interpretable prediction of fatigue properties of AM materials remains to be explored.

\textbf{\textit{Unable extension} from AM material fatigue data to structure life design by ML models.} Currently, most of the ML models are restricted to the prediction of fatigue properties of standard specimens fabricated by AM techniques and various post-processing treatments. For the real application of AM materials to structures that endure cyclic loadings and has the risk of fatigue failure, how the standard fatigue data from AM materials is extended to the design of fatigue life, damage tolerance and durability of structures by ML models remains as an issue to be explored. This issue is critical for replacing aerospace structures manufactured by traditional techniques with light and net-shape AM materials. Especially for the damage tolerance design of aerospace structures fabricated by AM techniques, the massive data on the fatigue crack growth rate of AM materials are compulsory, but are currently not as ready as that of materials manufactured by traditional techniques. Integrating ML strategies, experience and criteria on structure life design, stress/strain analysis by numerical simulations, and newly developed methodologies for crack propagation could be a route to extend AM material fatigue data to structure life design.

\section*{CRediT authorship contribution statement}
\textbf{Min Yi:} Conceptualization, Investigation, Resources, Supervision, Project administration, Funding acquisition, Writing -- original draft \& review \& editing. \textbf{Ming Xue:} Conceptualization, Investigation, Data curation, Writing -- original draft \& review \& editing. \textbf{Peihong Cong:} Conceptualization, Project administration, Funding acquisition, Writing -- review. \textbf{Yang Song:} Conceptualization, Project administration, Funding acquisition, Writing -- review. \textbf{Haiyang Zhang:} Conceptualization, Investigation, Data curation, Writing -- review. \textbf{Lingfeng Wang:} Conceptualization, Investigation, Data curation, Writing -- original draft. \textbf{Liucheng Zhou:} Conceptualization, Project administration, Funding acquisition, Writing -- review.  \textbf{Yinghong Li:} Conceptualization, Project administration, Funding acquisition, Writing -- review.
\textbf{Wanlin Guo:} Conceptualization, Project administration, Funding acquisition, Writing -- review.

\section*{Declaration of Competing Interest}
The authors declare that they have no known competing financial interests or personal relationships that could have appeared to influence the work reported in this paper.

\section*{Acknowledgements}
The authors acknowledge the support from National Science and Technology Major Project (J2019-IV-0014-0082), National Key Research and Development Program of China (2022YFB4600700), National Overseas Youth Talents Program, the Research Fund of State Key Laboratory of Mechanics and Control for Aerospace Structures, and a project Funded by the Priority Academic Program Development of Jiangsu Higher Education Institutions. This work is partially supported by High Performance Computing Platform of Nanjing University of Aeronautics and Astronautics. Simulations were also performed on Hefei advanced computing center.

\section*{Data Availability}
Data will be made available on request.


\bibliography{xuebibfile}

\begin{thebibliography}{10}
\expandafter\ifx\csname url\endcsname\relax
  \def\url#1{\texttt{#1}}\fi
\expandafter\ifx\csname urlprefix\endcsname\relax\def\urlprefix{URL }\fi
\expandafter\ifx\csname href\endcsname\relax
  \def\href#1#2{#2} \def\path#1{#1}\fi

\bibitem{Park2022}
S.~Park, W.~Shou, L.~Makatura, W.~Matusik, K.~K. Fu, {3D printing of polymer
  composites: Materials, processes, and applications}, Matter 5~(1) (2022)
  43--76.
\newblock \href {http://dx.doi.org/10.1016/J.MATT.2021.10.018}
  {\path{doi:10.1016/J.MATT.2021.10.018}}.

\bibitem{Herzog2016}
D.~Herzog, V.~Seyda, E.~Wycisk, C.~Emmelmann, {Additive manufacturing of
  metals}, Acta Materialia 117 (2016) 371--392.
\newblock \href {http://dx.doi.org/10.1016/J.ACTAMAT.2016.07.019}
  {\path{doi:10.1016/J.ACTAMAT.2016.07.019}}.

\bibitem{Zhang2019}
D.~Zhang, D.~Qiu, M.~A. Gibson, Y.~Zheng, H.~L. Fraser, D.~H. StJohn, M.~A.
  Easton, {Additive manufacturing of ultrafine-grained high-strength titanium
  alloys}, Nature 576~(7785) (2019) 91--95.
\newblock \href {http://dx.doi.org/10.1038/s41586-019-1783-1}
  {\path{doi:10.1038/s41586-019-1783-1}}.

\bibitem{Yi2023Modeling}
M.~Yi, W.~Wang, M.~Xue, Q.~Gong, B.-X. Xu, {Modeling and simulation of
  sintering process across scales}, Archives of Computational Methods in
  Engineering 30 (2023) 1--34.
\newblock \href {http://dx.doi.org/10.1007/s11831-023-09905-0}
  {\path{doi:10.1007/s11831-023-09905-0}}.

\bibitem{Benedetti2021}
M.~Benedetti, A.~du~Plessis, R.~O. Ritchie, M.~Dallago, S.~M. Razavi, F.~Berto,
  {Architected cellular materials: A review on their mechanical properties
  towards fatigue-tolerant design and fabrication}, Materials Science and
  Engineering: R: Reports 144 (2021) 100606.
\newblock \href {http://dx.doi.org/10.1016/J.MSER.2021.100606}
  {\path{doi:10.1016/J.MSER.2021.100606}}.

\bibitem{LI2023456Collaborative}
S.~Li, H.~Wei, S.~Yuan, J.~Zhu, J.~Li, W.~Zhang, Collaborative optimization
  design of process parameter and structural topology for laser additive
  manufacturing, Chinese Journal of Aeronautics 36~(1) (2023) 456--467.
\newblock \href {http://dx.doi.org/10.1016/j.cja.2021.12.010}
  {\path{doi:10.1016/j.cja.2021.12.010}}.

\bibitem{Culmone2019}
C.~Culmone, G.~Smit, P.~Breedveld, {Additive manufacturing of medical
  instruments: A state-of-the-art review}, Additive Manufacturing 27 (2019)
  461--473.
\newblock \href {http://dx.doi.org/10.1016/J.ADDMA.2019.03.015}
  {\path{doi:10.1016/J.ADDMA.2019.03.015}}.

\bibitem{ZHU202191Areview}
J.~Zhu, H.~Zhou, C.~Wang, L.~Zhou, S.~Yuan, W.~Zhang, A review of topology
  optimization for additive manufacturing: Status and challenges, Chinese
  Journal of Aeronautics 34~(1) (2021) 91--110.
\newblock \href {http://dx.doi.org/10.1016/j.cja.2020.09.020}
  {\path{doi:10.1016/j.cja.2020.09.020}}.

\bibitem{Zhakeyev2017}
A.~Zhakeyev, P.~Wang, L.~Zhang, W.~Shu, H.~Wang, J.~Xuan, A.~Zhakeyev, H.~Wang,
  J.~Xuan, P.~Wang, L.~Zhang, W.~Shu, {Additive Manufacturing: Unlocking the
  Evolution of Energy Materials}, Advanced Science 4~(10) (2017) 1700187.
\newblock \href {http://dx.doi.org/10.1002/ADVS.201700187}
  {\path{doi:10.1002/ADVS.201700187}}.

\bibitem{SHI20201252Anaerospace}
G.~Shi, C.~Guan, D.~Quan, D.~Wu, L.~Tang, T.~Gao, An aerospace bracket designed
  by thermo-elastic topology optimization and manufactured by additive
  manufacturing, Chinese Journal of Aeronautics 33~(4) (2020) 1252--1259.
\newblock \href {http://dx.doi.org/10.1016/j.cja.2019.09.006}
  {\path{doi:10.1016/j.cja.2019.09.006}}.

\bibitem{DuPlessis2022}
A.~du~Plessis, S.~M.~J. Razavi, M.~Benedetti, S.~Murchio, M.~Leary, M.~Watson,
  D.~Bhate, F.~Berto, {Properties and applications of additively manufactured
  metallic cellular materials: A review}, Progress in Materials Science 125
  (2022) 100918.
\newblock \href {http://dx.doi.org/10.1016/J.PMATSCI.2021.100918}
  {\path{doi:10.1016/J.PMATSCI.2021.100918}}.

\bibitem{ZHOU20191727Lightweight}
H.~Zhou, X.~Zhang, H.~Zeng, H.~Yang, H.~Lei, X.~Li, Y.~Wang, Lightweight
  structure of a phase-change thermal controller based on lattice cells
  manufactured by {SLM}, Chinese Journal of Aeronautics 32~(7) (2019)
  1727--1732.
\newblock \href {http://dx.doi.org/10.1016/j.cja.2018.08.017}
  {\path{doi:10.1016/j.cja.2018.08.017}}.

\bibitem{Yi2019Computational}
M.~Yi, B.-X. Xu, O.~Gutfleisch, {Computational study on microstructure
  evolution and magnetic property of laser additively manufactured magnetic
  materials}, Computational Mechanics 64~(4) (2019) 917--935.
\newblock \href {http://dx.doi.org/10.1007/s00466-019-01687-2}
  {\path{doi:10.1007/s00466-019-01687-2}}.

\bibitem{Yang20193Dnon}
Y.~Yang, O.~Ragnvaldsen, Y.~Bai, M.~Yi, B.-X. Xu, {3D non-isothermal
  phase-field simulation of microstructure evolution during selective laser
  sintering}, npj Computational Materials 5 (2019) 81.
\newblock \href {http://dx.doi.org/10.1038/s41524-019-0219-7}
  {\path{doi:10.1038/s41524-019-0219-7}}.

\bibitem{WANG2020101538Machine}
C.~Wang, X.~Tan, S.~Tor, C.~Lim, Machine learning in additive manufacturing:
  State-of-the-art and perspectives, Additive Manufacturing 36 (2020) 101538.
\newblock \href {http://dx.doi.org/10.1016/j.addma.2020.101538}
  {\path{doi:10.1016/j.addma.2020.101538}}.

\bibitem{Pham2005}
D.~T. Pham, A.~A. Afify, {Machine-learning techniques and their applications in
  manufacturing}, Proceedings of the Institution of Mechanical Engineers, Part
  B: Journal of Engineering Manufacture 219~(5) (2005) 395--412.
\newblock \href {http://dx.doi.org/10.1243/095440505X32274}
  {\path{doi:10.1243/095440505X32274}}.

\bibitem{Voulodimos2018}
A.~Voulodimos, N.~Doulamis, A.~Doulamis, E.~Protopapadakis, {Deep Learning for
  Computer Vision: A Brief Review}, Computational Intelligence and Neuroscience
  2018.
\newblock \href {http://dx.doi.org/10.1155/2018/7068349}
  {\path{doi:10.1155/2018/7068349}}.

\bibitem{CAI2022107580}
R.~Cai, K.~Wang, W.~Wen, Y.~Peng, M.~Baniassadi, S.~Ahzi, Application of
  machine learning methods on dynamic strength analysis for additive
  manufactured polypropylene-based composites, Polymer Testing 110 (2022)
  107580.
\newblock \href {http://dx.doi.org/10.1016/j.polymertesting.2022.107580}
  {\path{doi:10.1016/j.polymertesting.2022.107580}}.

\bibitem{Mahadevkar2022}
S.~V. Mahadevkar, B.~Khemani, S.~Patil, K.~Kotecha, D.~R. Vora, A.~Abraham,
  L.~A. Gabralla, {A Review on Machine Learning Styles in Computer Vision -
  Techniques and Future Directions}, IEEE Access 10~(99) (2022) 107293--107329.
\newblock \href {http://dx.doi.org/10.1109/ACCESS.2022.3209825}
  {\path{doi:10.1109/ACCESS.2022.3209825}}.

\bibitem{Ajani2021}
T.~S. Ajani, A.~L. Imoize, A.~A. Atayero, {An overview of machine learning
  within embedded and mobile devices-optimizations and applications}, Sensors
  21~(13).
\newblock \href {http://dx.doi.org/10.3390/s21134412}
  {\path{doi:10.3390/s21134412}}.

\bibitem{Kaul2022}
A.~Kaul, S.~Raina, {Support vector machine versus convolutional neural network
  for hyperspectral image classification: A systematic review}, Concurrency and
  Computation: Practice and Experience 34~(15) (2022) e6945.
\newblock \href {http://dx.doi.org/10.1002/cpe.6945}
  {\path{doi:10.1002/cpe.6945}}.

\bibitem{Sanjana2021}
S.~Sanjana, S.~Sanjana, V.~R. Shriya, G.~Vaishnavi, K.~Ashwini, {A review on
  various methodologies used for vehicle classification, helmet detection and
  number plate recognition}, Evolutionary Intelligence 14~(2) (2021) 979--987.
\newblock \href {http://dx.doi.org/10.1007/s12065-020-00493-7}
  {\path{doi:10.1007/s12065-020-00493-7}}.

\bibitem{Mozaffari2022}
S.~Mozaffari, O.~Y. Al-Jarrah, M.~Dianati, P.~Jennings, A.~Mouzakitis, {Deep
  Learning-Based Vehicle Behavior Prediction for Autonomous Driving
  Applications: A Review}, IEEE Transactions on Intelligent Transportation
  Systems 23~(1) (2022) 33--47.
\newblock \href {http://dx.doi.org/10.1109/TITS.2020.3012034}
  {\path{doi:10.1109/TITS.2020.3012034}}.

\bibitem{Mazhari2021}
A.~A. Mazhari, R.~Ticknor, S.~Swei, S.~Krzesniak, M.~Teodorescu, {Automated
  Testing and Characterization of Additive Manufacturing (ATCAM)}, Journal of
  Materials Engineering and Performance 30~(9) (2021) 6862--6873.
\newblock \href {http://dx.doi.org/10.1007/S11665-021-06042-2/FIGURES/8}
  {\path{doi:10.1007/S11665-021-06042-2/FIGURES/8}}.

\bibitem{Rovinelli2018}
A.~Rovinelli, M.~D. Sangid, H.~Proudhon, W.~Ludwig, Using machine learning and
  a data-driven approach to identify the small fatigue crack driving force in
  polycrystalline materials, npj Computational Materials 4~(1) (2018) 35.
\newblock \href {http://dx.doi.org/10.1038/s41524-018-0094-7}
  {\path{doi:10.1038/s41524-018-0094-7}}.

\bibitem{WAN20191137}
H.~Wan, G.~Chen, C.~Li, X.~Qi, G.~Zhang, Data-driven evaluation of fatigue
  performance of additive manufactured parts using miniature specimens, Journal
  of Materials Science \& Technology 35~(6) (2019) 1137--1146.
\newblock \href {http://dx.doi.org/10.1016/j.jmst.2018.12.011}
  {\path{doi:10.1016/j.jmst.2018.12.011}}.

\bibitem{ZHANG2022106808}
J.~Zhang, J.~Zhu, W.~Guo, W.~Guo, A machine learning-based approach to predict
  the fatigue life of three-dimensional cracked specimens, International
  Journal of Fatigue 159 (2022) 106808.
\newblock \href {http://dx.doi.org/10.1016/j.ijfatigue.2022.106808}
  {\path{doi:10.1016/j.ijfatigue.2022.106808}}.

\bibitem{Zhou13532}
K.~Zhou, X.~Sun, S.~Shi, K.~Song, X.~Chen, Machine learning-based genetic
  feature identification and fatigue life prediction, Fatigue \& Fracture of
  Engineering Materials \& Structures 44~(9) (2021) 2524--2537.
\newblock \href {http://dx.doi.org/10.1111/ffe.13532}
  {\path{doi:10.1111/ffe.13532}}.

\bibitem{GORJI2022106949}
M.~B. Gorji, A.~{de Pannemaecker}, S.~Spevack, Machine learning predicts
  fretting and fatigue key mechanical properties, International Journal of
  Mechanical Sciences 215 (2022) 106949.
\newblock \href {http://dx.doi.org/10.1016/j.ijmecsci.2021.106949}
  {\path{doi:10.1016/j.ijmecsci.2021.106949}}.

\bibitem{Nasiri2021}
S.~Nasiri, M.~R. Khosravani, {Machine learning in predicting mechanical
  behavior of additively manufactured parts}, Journal of Materials Research and
  Technology 14 (2021) 1137--1153.
\newblock \href {http://dx.doi.org/10.1016/j.jmrt.2021.07.004}
  {\path{doi:10.1016/j.jmrt.2021.07.004}}.

\bibitem{Guo2022}
S.~Guo, M.~Agarwal, C.~Cooper, Q.~Tian, R.~X. Gao, W.~G. Guo, Y.~B. Guo,
  {Machine learning for metal additive manufacturing: Towards a
  physics-informed data-driven paradigm}, Journal of Manufacturing Systems
  62~(November 2021) (2022) 145--163.
\newblock \href {http://dx.doi.org/10.1016/j.jmsy.2021.11.003}
  {\path{doi:10.1016/j.jmsy.2021.11.003}}.

\bibitem{Baumann2018}
F.~W. Baumann, A.~Sekulla, M.~Hassler, B.~Himpel, M.~Pfeil, {Trends of machine
  learning in additive manufacturing}, International Journal of Rapid
  Manufacturing 7~(4) (2018) 310.
\newblock \href {http://dx.doi.org/10.1504/ijrapidm.2018.10016883}
  {\path{doi:10.1504/ijrapidm.2018.10016883}}.

\bibitem{Meng2020Machine}
L.~Meng, B.~McWilliams, W.~Jarosinski, H.-Y. Park, Y.-G. Jung, J.~Lee,
  J.~Zhang, {Machine Learning in Additive Manufacturing: A Review}, JOM 72~(6)
  (2020) 2363--2377.
\newblock \href {http://dx.doi.org/10.1007/s11837-020-04155-y}
  {\path{doi:10.1007/s11837-020-04155-y}}.

\bibitem{Grierson2021Machine}
D.~Grierson, A.~E.~W. Rennie, S.~D. Quayle, Machine learning for additive
  manufacturing, Encyclopedia 1~(3) (2021) 576--588.
\newblock \href {http://dx.doi.org/10.3390/encyclopedia1030048}
  {\path{doi:10.3390/encyclopedia1030048}}.

\bibitem{10.1111/ffe.13640}
J.~Chen, Y.~Liu, Fatigue modeling using neural networks: A comprehensive
  review, Fatigue \& Fracture of Engineering Materials \& Structures 45~(4)
  (2022) 945--979.
\newblock \href {http://dx.doi.org/10.1111/ffe.13640}
  {\path{doi:10.1111/ffe.13640}}.

\bibitem{Ladani2021}
L.~J. Ladani, {Applications of artificial intelligence and machine learning in
  metal additive manufacturing}, JPhys Materials 4~(4).
\newblock \href {http://dx.doi.org/10.1088/2515-7639/ac2791}
  {\path{doi:10.1088/2515-7639/ac2791}}.

\bibitem{JIN20201541Machine}
Z.~Jin, Z.~Zhang, K.~Demir, G.~X. Gu, Machine learning for advanced additive
  manufacturing, Matter 3~(5) (2020) 1541--1556.
\newblock \href {http://dx.doi.org/10.1016/j.matt.2020.08.023}
  {\path{doi:10.1016/j.matt.2020.08.023}}.

\bibitem{QI2019721}
X.~Qi, G.~Chen, Y.~Li, X.~Cheng, C.~Li, Applying neural-network-based machine
  learning to additive manufacturing: Current applications, challenges, and
  future perspectives, Engineering 5~(4) (2019) 721--729.
\newblock \href {http://dx.doi.org/10.1016/j.eng.2019.04.012}
  {\path{doi:10.1016/j.eng.2019.04.012}}.

\bibitem{Kumar2022Machine}
S.~Kumar, T.~Gopi, N.~Harikeerthana, M.~K. Gupta, V.~Gaur, G.~M. Krolczyk,
  C.~Wu, {Machine learning techniques in additive manufacturing: a state of the
  art review on design, processes and production control}, Journal of
  Intelligent Manufacturing 34 (2023) 21--55.
\newblock \href {http://dx.doi.org/10.1007/s10845-022-02029-5}
  {\path{doi:10.1007/s10845-022-02029-5}}.

\bibitem{Raza2022Incorporation}
A.~Raza, K.~M. Deen, R.~Jaafreh, K.~Hamad, A.~Haider, W.~Haider, {Incorporation
  of machine learning in additive manufacturing: a review}, The International
  Journal of Advanced Manufacturing Technology 122~(3-4) (2022) 1143--1166.
\newblock \href {http://dx.doi.org/10.1007/s00170-022-09916-4}
  {\path{doi:10.1007/s00170-022-09916-4}}.

\bibitem{Qin2022}
J.~Qin, F.~Hu, Y.~Liu, P.~Witherell, C.~C. Wang, D.~W. Rosen, T.~W. Simpson,
  Y.~Lu, Q.~Tang, {Research and application of machine learning for additive
  manufacturing}, Additive Manufacturing 52 (2022) 102691.
\newblock \href {http://dx.doi.org/10.1016/j.addma.2022.102691}
  {\path{doi:10.1016/j.addma.2022.102691}}.

\bibitem{SCHMIDHUBER201585}
J.~Schmidhuber, Deep learning in neural networks: An overview, Neural Networks
  61 (2015) 85--117.
\newblock \href {http://dx.doi.org/10.1016/j.neunet.2014.09.003}
  {\path{doi:10.1016/j.neunet.2014.09.003}}.

\bibitem{XU201018}
H.~Xu, B.~Yu, Automatic thesaurus construction for spam filtering using revised
  back propagation neural network, Expert Systems with Applications 37~(1)
  (2010) 18--23.
\newblock \href {http://dx.doi.org/10.1016/j.eswa.2009.02.059}
  {\path{doi:10.1016/j.eswa.2009.02.059}}.

\bibitem{MUHAMMAD2021102867}
W.~Muhammad, A.~P. Brahme, O.~Ibragimova, J.~Kang, K.~Inal, A machine learning
  framework to predict local strain distribution and the evolution of plastic
  anisotropy and fracture in additively manufactured alloys, International
  Journal of Plasticity 136 (2021) 102867.
\newblock \href {http://dx.doi.org/10.1016/j.ijplas.2020.102867}
  {\path{doi:10.1016/j.ijplas.2020.102867}}.

\bibitem{LI20031861}
H.-X. Li, E.~Lee, Interpolation functions of feedforward neural networks,
  Computers \& Mathematics with Applications 46~(12) (2003) 1861--1874.
\newblock \href {http://dx.doi.org/10.1016/S0898-1221(03)90242-2}
  {\path{doi:10.1016/S0898-1221(03)90242-2}}.

\bibitem{lan2017conditional}
G.~Lan, S.~Pokutta, Y.~Zhou, D.~Zink, Conditional accelerated lazy stochastic
  gradient descent (2017).
\newblock \href {http://arxiv.org/abs/1703.05840} {\path{arXiv:1703.05840}}.

\bibitem{LIU2019129}
Z.~Liu, Y.~Cao, Y.~Wang, W.~Wang, Computer vision-based concrete crack
  detection using u-net fully convolutional networks, Automation in
  Construction 104 (2019) 129--139.
\newblock \href {http://dx.doi.org/10.1016/j.autcon.2019.04.005}
  {\path{doi:10.1016/j.autcon.2019.04.005}}.

\bibitem{ZHAN2022108352}
Z.~Zhan, N.~Ao, Y.~Hu, C.~Liu, Defect‐induced fatigue scattering and
  assessment of additively manufactured {300M-AerMet100} steel: An
  investigation based on experiments and machine learning, Engineering Fracture
  Mechanics 264 (2022) 108352.
\newblock \href {http://dx.doi.org/10.1016/j.engfracmech.2022.108352}
  {\path{doi:10.1016/j.engfracmech.2022.108352}}.

\bibitem{LIU2022106836}
S.~Liu, W.~Shi, Z.~Zhan, W.~Hu, Q.~Meng, On the development of error-trained
  {BP-ANN technique with CDM model for the HCF} life prediction of aluminum
  alloy, International Journal of Fatigue 160 (2022) 106836.
\newblock \href {http://dx.doi.org/10.1016/j.ijfatigue.2022.106836}
  {\path{doi:10.1016/j.ijfatigue.2022.106836}}.

\bibitem{ZHAN2021106089}
Z.~Zhan, H.~Li, A novel approach based on the elastoplastic fatigue damage and
  machine learning models for life prediction of aerospace alloy parts
  fabricated by additive manufacturing, International Journal of Fatigue 145
  (2021) 106089.
\newblock \href {http://dx.doi.org/10.1016/j.ijfatigue.2020.106089}
  {\path{doi:10.1016/j.ijfatigue.2020.106089}}.

\bibitem{SNOW2022117476}
Z.~Snow, E.~W. Reutzel, J.~Petrich, Correlating in-situ sensor data to defect
  locations and part quality for additively manufactured parts using machine
  learning, Journal of Materials Processing Technology 302 (2022) 117476.
\newblock \href {http://dx.doi.org/10.1016/j.jmatprotec.2021.117476}
  {\path{doi:10.1016/j.jmatprotec.2021.117476}}.

\bibitem{Vaz2021}
J.~M. Vaz, S.~Balaji, Convolutional neural networks {(CNNs)}: concepts and
  applications in pharmacogenomics, Molecular diversity 25~(3) (2021)
  1569--1584.
\newblock \href {http://dx.doi.org/10.1007/s11030-021-10225-3}
  {\path{doi:10.1007/s11030-021-10225-3}}.

\bibitem{7426826}
N.~Tajbakhsh, J.~Y. Shin, S.~R. Gurudu, R.~T. Hurst, C.~B. Kendall, M.~B.
  Gotway, J.~Liang, Convolutional neural networks for medical image analysis:
  Full training or fine tuning?, {IEEE} Transactions on Medical Imaging 35~(5)
  (2016) 1299--1312.
\newblock \href {http://dx.doi.org/10.1109/TMI.2016.2535302}
  {\path{doi:10.1109/TMI.2016.2535302}}.

\bibitem{LeCun2015}
Y.~LeCun, Y.~Bengio, G.~Hinton, Deep learning, Nature 521~(7553) (2015)
  436--444.
\newblock \href {http://dx.doi.org/10.1038/nature14539}
  {\path{doi:10.1038/nature14539}}.

\bibitem{LIMAJUNIOR2020106191}
F.~R. Lima-Junior, L.~C.~R. Carpinetti, An adaptive network-based fuzzy
  inference system to supply chain performance evaluation based on {SCOR}
  metrics, Computers \& Industrial Engineering 139 (2020) 106191.
\newblock \href {http://dx.doi.org/10.1016/j.cie.2019.106191}
  {\path{doi:10.1016/j.cie.2019.106191}}.

\bibitem{256541}
J.-S. Jang, {ANFIS}: adaptive-network-based fuzzy inference system, {IEEE}
  Transactions on Systems, Man, and Cybernetics 23~(3) (1993) 665--685.
\newblock \href {http://dx.doi.org/10.1109/21.256541}
  {\path{doi:10.1109/21.256541}}.

\bibitem{Aengchuan2018}
P.~Aengchuan, B.~Phruksaphanrat, Comparison of fuzzy inference system {(FIS),
  FIS with artificial neural networks (FIS + ANN) and FIS with adaptive
  neuro-fuzzy inference system (FIS + ANFIS)} for inventory control, Journal of
  Intelligent Manufacturing 29~(4) (2018) 905--923.
\newblock \href {http://dx.doi.org/10.1007/s10845-015-1146-1}
  {\path{doi:10.1007/s10845-015-1146-1}}.

\bibitem{TAKAGI198355}
T.~Takagi, M.~Sugeno, Derivation of fuzzy control rules from human operator's
  control actions, {IFAC} Proceedings Volumes 16~(13) (1983) 55--60.
\newblock \href {http://dx.doi.org/10.1016/S1474-6670(17)62005-6}
  {\path{doi:10.1016/S1474-6670(17)62005-6}}.

\bibitem{ZHANG2019105194}
M.~Zhang, C.-N. Sun, X.~Zhang, P.~C. Goh, J.~Wei, D.~Hardacre, H.~Li, High
  cycle fatigue life prediction of laser additive manufactured stainless steel:
  A machine learning approach, International Journal of Fatigue 128 (2019)
  105194.
\newblock \href {http://dx.doi.org/10.1016/j.ijfatigue.2019.105194}
  {\path{doi:10.1016/j.ijfatigue.2019.105194}}.

\bibitem{BELFIORE20071705}
N.~Belfiore, F.~Ianniello, D.~Stocchi, F.~Casadei, D.~Bazzoni, A.~Finzi,
  S.~Carrara, J.~Gonzalez, J.~Llanos, I.~Heikkila, F.~Penalba, X.~Gomez, A
  hybrid approach to the development of a multilayer neural network for wear
  and fatigue prediction in metal forming, Tribology International 40~(10)
  (2007) 1705--1717.
\newblock \href {http://dx.doi.org/10.1016/j.triboint.2007.01.008}
  {\path{doi:10.1016/j.triboint.2007.01.008}}.

\bibitem{cuomo2022scientific}
S.~Cuomo, V.~S. di~Cola, F.~Giampaolo, G.~Rozza, M.~Raissi, F.~Piccialli,
  Scientific machine learning through physics-informed neural networks: Where
  we are and what's next (2022).
\newblock \href {http://arxiv.org/abs/2201.05624} {\path{arXiv:2201.05624}}.

\bibitem{Karniadakis2021}
G.~E. Karniadakis, I.~G. Kevrekidis, L.~Lu, P.~Perdikaris, S.~Wang, L.~Yang,
  Physics-informed machine learning, Nature Reviews Physics 3~(6) (2021)
  422--440.
\newblock \href {http://dx.doi.org/10.1038/s42254-021-00314-5}
  {\path{doi:10.1038/s42254-021-00314-5}}.

\bibitem{202211}
S.~Wang, X.~Yu, P.~Perdikaris, When and why {PINNs} fail to train: A neural
  tangent kernel perspective, Journal of Computational Physics 449 (2022)
  110768.
\newblock \href {http://dx.doi.org/10.1016/j.jcp.2021.110768}
  {\path{doi:10.1016/j.jcp.2021.110768}}.

\bibitem{CHEN2021114316}
J.~Chen, Y.~Liu, Probabilistic physics-guided machine learning for fatigue data
  analysis, Expert Systems with Applications 168 (2021) 114316.
\newblock \href {http://dx.doi.org/10.1016/j.eswa.2020.114316}
  {\path{doi:10.1016/j.eswa.2020.114316}}.

\bibitem{CHEN2021101876}
J.~Chen, Y.~Liu, Fatigue property prediction of additively manufactured
  {Ti-6Al-4V} using probabilistic physics-guided learning, Additive
  Manufacturing 39 (2021) 101876.
\newblock \href {http://dx.doi.org/10.1016/j.addma.2021.101876}
  {\path{doi:10.1016/j.addma.2021.101876}}.

\bibitem{doi:10.1080/00401706.1999.10485925}
F.~G.Pascual, W.~Q. Meeker, Estimating fatigue curves with the random
  fatigue-limit model, Technometrics 41~(4) (1999) 277--289.
\newblock \href {http://dx.doi.org/10.1080/00401706.1999.10485925}
  {\path{doi:10.1080/00401706.1999.10485925}}.

\bibitem{Lopes2012357}
N.~Lopes, B.~Ribeiro, Handling missing values via a neural selective input
  model, Neural Network World 22~(4) (2012) 357 – 370.
\newblock \href {http://dx.doi.org/10.14311/NNW.2012.22.021}
  {\path{doi:10.14311/NNW.2012.22.021}}.

\bibitem{Noble2006}
W.~S. Noble, What is a support vector machine?, Nature Biotechnology 24~(12)
  (2006) 1565--1567.
\newblock \href {http://dx.doi.org/10.1038/nbt1206-1565}
  {\path{doi:10.1038/nbt1206-1565}}.

\bibitem{Cortes1995}
C.~Cortes, V.~Vapnik, Support-vector networks, Machine Learning 20~(3) (1995)
  273--297.
\newblock \href {http://dx.doi.org/10.1007/BF00994018}
  {\path{doi:10.1007/BF00994018}}.

\bibitem{SANCHEZA20035}
V.~{Sanchez A}, Advanced support vector machines and kernel methods,
  Neurocomputing 55~(1--2) (2003) 5--20.
\newblock \href {http://dx.doi.org/10.1016/S0925-2312(03)00373-4}
  {\path{doi:10.1016/S0925-2312(03)00373-4}}.

\bibitem{DANG2022106748}
L.~Dang, X.~He, D.~Tang, Y.~Li, T.~Wang, A fatigue life prediction approach for
  laser-directed energy deposition titanium alloys by using support vector
  regression based on pore-induced failures, International Journal of Fatigue
  159 (2022) 106748.
\newblock \href {http://dx.doi.org/10.1016/j.ijfatigue.2022.106748}
  {\path{doi:10.1016/j.ijfatigue.2022.106748}}.

\bibitem{LUO2021140693}
Y.~Luo, B.~Zhang, X.~Feng, Z.~Song, X.~Qi, C.~Li, G.~Chen, G.~Zhang,
  Pore-affected fatigue life scattering and prediction of additively
  manufactured inconel 718: An investigation based on miniature specimen
  testing and machine learning approach, Materials Science and Engineering: A
  802 (2021) 140693.
\newblock \href {http://dx.doi.org/10.1016/j.msea.2020.140693}
  {\path{doi:10.1016/j.msea.2020.140693}}.

\bibitem{WANG2019373}
X.~Wang, X.~He, T.~Wang, Y.~Li, Internal pores in {DED Ti-6.5Al-2Zr-Mo-V} alloy
  and their influence on crack initiation and fatigue life in the mid-life
  regime, Additive Manufacturing 28 (2019) 373--393.
\newblock \href {http://dx.doi.org/10.1016/j.addma.2019.05.007}
  {\path{doi:10.1016/j.addma.2019.05.007}}.

\bibitem{SHERIDAN2021106033}
L.~Sheridan, J.~E. Gockel, O.~E. Scott-Emuakpor, Stress-defect-life
  interactions of fatigued additively manufactured alloy 718, International
  Journal of Fatigue 143 (2021) 106033.
\newblock \href {http://dx.doi.org/10.1016/j.ijfatigue.2020.106033}
  {\path{doi:10.1016/j.ijfatigue.2020.106033}}.

\bibitem{ZHAN2021105941}
Z.~Zhan, H.~Li, Machine learning based fatigue life prediction with effects of
  additive manufacturing process parameters for printed {SS 316L},
  International Journal of Fatigue 142 (2021) 105941.
\newblock \href {http://dx.doi.org/10.1016/j.ijfatigue.2020.105941}
  {\path{doi:10.1016/j.ijfatigue.2020.105941}}.

\bibitem{LI2022107018}
A.~Li, S.~Baig, J.~Liu, S.~Shao, N.~Shamsaei, Defect criticality analysis on
  fatigue life of {L-PBF 17-4 PH} stainless steel via machine learning,
  International Journal of Fatigue 163 (2022) 107018.
\newblock \href {http://dx.doi.org/10.1016/j.ijfatigue.2022.107018}
  {\path{doi:10.1016/j.ijfatigue.2022.107018}}.

\bibitem{BAO2021107508}
H.~Bao, S.~Wu, Z.~Wu, G.~Kang, X.~Peng, P.~J. Withers, A machine-learning
  fatigue life prediction approach of additively manufactured metals,
  Engineering Fracture Mechanics 242 (2021) 107508.
\newblock \href {http://dx.doi.org/10.1016/j.engfracmech.2020.107508}
  {\path{doi:10.1016/j.engfracmech.2020.107508}}.

\bibitem{PENG2022107185}
X.~Peng, S.~Wu, W.~Qian, J.~Bao, Y.~Hu, Z.~Zhan, G.~Guo, P.~J. Withers, The
  potency of defects on fatigue of additively manufactured metals,
  International Journal of Mechanical Sciences 221 (2022) 107185.
\newblock \href {http://dx.doi.org/10.1016/j.ijmecsci.2022.107185}
  {\path{doi:10.1016/j.ijmecsci.2022.107185}}.

\bibitem{ZHAN2021107850}
Z.~Zhan, W.~Hu, Q.~Meng, Data-driven fatigue life prediction in additive
  manufactured titanium alloy: A damage mechanics based machine learning
  framework, Engineering Fracture Mechanics 252 (2021) 107850.
\newblock \href {http://dx.doi.org/10.1016/j.engfracmech.2021.107850}
  {\path{doi:10.1016/j.engfracmech.2021.107850}}.

\bibitem{Breiman2001}
L.~Breiman, Random forests, Machine Learning 45~(1) (2001) 5--32.
\newblock \href {http://dx.doi.org/10.1023/A:1010933404324}
  {\path{doi:10.1023/A:1010933404324}}.

\bibitem{WANG2022107147}
H.~Wang, B.~Li, F.-Z. Xuan, Fatigue-life prediction of additively manufactured
  metals by continuous damage mechanics {(CDM)}-informed machine learning with
  sensitive features, International Journal of Fatigue 164 (2022) 107147.
\newblock \href {http://dx.doi.org/10.1016/j.ijfatigue.2022.107147}
  {\path{doi:10.1016/j.ijfatigue.2022.107147}}.

\bibitem{met12010050}
N.~Konda, R.~Verma, R.~Jayaganthan, Machine learning based predictions of
  fatigue crack growth rate of additively manufactured {Ti6Al4V}, Metals
  12~(1).
\newblock \href {http://dx.doi.org/10.3390/met12010050}
  {\path{doi:10.3390/met12010050}}.

\bibitem{WOS:000518708500028}
J.~Chen, S.~Liu, W.~Zhang, Y.~Liu, Uncertainty quantification of fatigue {S-N}
  curves with sparse data using hierarchical bayesian data augmentation,
  Interational Journal of Fatigue 134 (2020) 105511.
\newblock \href {http://dx.doi.org/10.1016/j.ijfatigue.2020.105511}
  {\path{doi:10.1016/j.ijfatigue.2020.105511}}.

\bibitem{doi:10.2514/6.2020-0680}
J.~Chen, Y.~Liu, {Uncertainty quantification of fatigue properties with sparse
  data using hierarchical Bayesian model}, AIAA Scitech 2020 Forum (2020) AIAA
  2020--0680. ~\href {http://dx.doi.org/10.2514/6.2020-0680}
  {\path{doi:10.2514/6.2020-0680}}.

\bibitem{WOS:000604573000113}
E.~Yang, Y.~Tang, L.~Li, W.~Yan, B.~Huang, Y.~Qiu, Research on the recurrent
  neural network-based fatigue damage model of asphalt binder and the finite
  element analysis development, Construction and Building Materials 267 (2021)
  121761.
\newblock \href {http://dx.doi.org/10.1016/j.conbuildmat.2020.121761}
  {\path{doi:10.1016/j.conbuildmat.2020.121761}}.

\bibitem{WOS:000074842400012}
Y.~Han, X.~Liu, S.~Dai, Fatigue life calculation of flawed structures based on
  artificial neural network with special learning set, International Journal of
  Pressure Vessels and Piping 75~(3) (1998) 263--269.
\newblock \href {http://dx.doi.org/10.1016/S0308-0161(98)00040-4}
  {\path{doi:10.1016/S0308-0161(98)00040-4}}.

\bibitem{kingma2013autoencoding}
D.~P. Kingma, M.~Welling, Auto-encoding variational bayes (2013).
\newblock \href {http://arxiv.org/abs/1312.6114} {\path{arXiv:1312.6114}}.

\bibitem{makhzani2015adversarial}
A.~Makhzani, J.~Shlens, N.~Jaitly, I.~Goodfellow, B.~Frey, Adversarial
  autoencoders (2015).
\newblock \href {http://arxiv.org/abs/1511.05644} {\path{arXiv:1511.05644}}.

\bibitem{Tang2023Modeling}
W.~Tang, Z.~Tang, W.~Lu, S.~Wang, M.~Yi, {Modeling and prediction of fatigue
  properties of additively manufactured metals}, Acta Mechanica Solida Sinica
  36 (2023) 181--213.
\newblock \href {http://dx.doi.org/10.1007/s10338-023-00380-5}
  {\path{doi:10.1007/s10338-023-00380-5}}.

\bibitem{Min2021Computational}
M.~Yi, K.~Chang, C.~Liang, L.~Zhou, Y.~Yang, X.~Yi, B.-X. Xu, {Computational
  study of evolution and fatigue dispersity of microstructures by additive
  manufacturing}, Chinese Journal of Theoretical and Applied Mechanic 53~(12)
  (2021) 3265--3275.
\newblock \href {http://dx.doi.org/10.6052/0459-1879-21-389}
  {\path{doi:10.6052/0459-1879-21-389}}.

\bibitem{1581138385}
A.~Y. Ng, Feature selection, {L1} vs. {L2} regularization, and rotational
  invariance, in: Proceedings of the Twenty-First International Conference on
  Machine Learning, ICML '04, Association for Computing Machinery, New York,
  NY, USA, 2004, p.~78.
\newblock \href {http://dx.doi.org/10.1145/1015330.1015435}
  {\path{doi:10.1145/1015330.1015435}}.

\bibitem{e23121673}
A.~Mohammad-Djafari, Regularization, bayesian inference, and machine learning
  methods for inverse problems, Entropy 23~(12).
\newblock \href {http://dx.doi.org/10.3390/e23121673}
  {\path{doi:10.3390/e23121673}}.

\bibitem{8402232}
H.~Li, H.~Zhao, H.~Li, Neural-response-based extreme learning machine for image
  classification, IEEE Transactions on Neural Networks and Learning Systems
  30~(2) (2019) 539--552.
\newblock \href {http://dx.doi.org/10.1109/TNNLS.2018.2845857}
  {\path{doi:10.1109/TNNLS.2018.2845857}}.

\bibitem{PRECHELT1998761}
L.~Prechelt, Automatic early stopping using cross validation: quantifying the
  criteria, Neural Networks 11~(4) (1998) 761--767.
\newblock \href {http://dx.doi.org/10.1016/S0893-6080(98)00010-0}
  {\path{doi:10.1016/S0893-6080(98)00010-0}}.

\bibitem{song2019does}
H.~Song, M.~Kim, D.~Park, J.-G. Lee, How does early stopping help
  generalization against label noise? (2019).
\newblock \href {http://arxiv.org/abs/1911.08059} {\path{arXiv:1911.08059}}.

\bibitem{XieSEPT1}
J.~Y. Xie, Z.~Y. Ma, J.~J. Lei, G.~Q. Zhang, J.~H. Xue, Z.~H. Tan, J.~Guo,
  Advanced dropout: A model-free methodology for bayesian dropout optimization,
  IEEE Transactions on Pattern Analysis and Machine Intelligence 44~(9) (2021)
  4605--4625.
\newblock \href {http://dx.doi.org/10.1109/TPAMI.2021.3083089}
  {\path{doi:10.1109/TPAMI.2021.3083089}}.

\end{thebibliography}

\end{document}